# Towards a Symbolic-Numeric Method to Compute Puiseux Series : The Modular Part

Adrien Poteaux  Marc Rybowicz

*XLIM-DMI (UMR CNRS 6172)*
*Université de Limoges*
*123 Avenue Albert Thomas*
*87060 Limoges - France*

**Abstract**

We have designed a new symbolic-numeric strategy to compute efficiently and accurately floating point Puiseux series defined by a bivariate polynomial over an algebraic number field. In essence, computations modulo a well chosen prime $p$ are used to obtain the exact information required to guide floating point computations. In this paper, we detail the symbolic part of our algorithm : First of all, we study modular reduction of Puiseux series and give a good reduction criterion to ensure that the information required by the numerical part is preserved. To establish our results, we introduce a simple modification of classical Newton polygons, that we call "generic Newton polygons", which happen to be very convenient. Then, we estimate the arithmetic complexity of computing Puiseux series over finite fields and improve known bounds. Finally, we give bit-complexity bounds for deterministic and randomized versions of the symbolic part. The details of the numerical part will be described in a forthcoming paper. The reader is referred for now to a preliminary version sketched in (Poteaux (2007)).

## 1. Introduction

Let $K$ be a number field and $F(X, Y)$ be a bivariate polynomial in $K[X, Y]$ such that :
- $F(X, Y) = Y^d + A_{d-1}(X)Y^{d-1} + \cdots + A_0(X)$, with $d > 0$ and $n = \deg_X F > 0$. We denote by $D$ the total degree of $F$.
- $F$ is squarefree.
- $\Delta_F(X)$ denotes the discriminant of $F$ with respect to $Y$.

*Email addresses:* `adrien.poteaux@xlim.fr` (Adrien Poteaux), `marc.rybowicz@xlim.fr` (Marc Rybowicz).

A root of $\Delta_F$ will be called a *critical point*. Critical points can also be defined as the set of numbers $x_0$ such that $F(x_0, Y)$ is not squarefree. The equation $F(X, Y) = 0$ defines $d$ algebraic functions of the variable $X$, which are analytic in any simply connected domain $\mathcal{D} \subset \mathbb{C}$ free of critical points.

Values of these algebraic functions can be computed by means of analytic continuation (see Chudnovsky and Chudnovsky (1986), Chudnovsky and Chudnovsky (1987), van der Hoeven (1999), van der Hoeven (2005), who require that a differential equation satisfied by the functions is known). In our context (see below), no differential equations is known a priori and we do not need a high precision since we just need enough information to separate functions. Hence, it is not clear that these asymptotically fast (in terms of precision) methods are relevant.

However, if $\mathcal{D}$ is included in a sufficiently small disc centered at a critical point $x_0$, it is well-known that numerical values of these functions in $\mathcal{D}$ can be obtained directly via truncated Puiseux series at $X = x_0$ (see Section 2).

We have used this fact to devise an algorithm to compute the monodromy of the Riemann sphere covering defined by the curve $F(X, Y) = 0$ (Poteaux (2007)). The algorithm follows paths along a minimal spanning tree for the set of critical points and expansions above critical point are used to bypass them.

In (Deconinck and van Hoeij (2001)), a different strategy for computing the monodromy is described. The approach therein and the implementation suffer from numerical instability and a lack of error control. These weaknesses have stimulated us for investigating the method mentioned above, that could hopefully offer a better numerical control. Another application of Puiseux series is the computation of Abel's map, at the heart of the celebrated Abel-Jacobi theorem (see for instance Miranda (1995) or Forster (1981)). Abel's map yields particularly interesting solutions of certain differential equations from Physics (see Deconinck and Patterson (2007) and Deconinck and Segur (1998)), and allows to answer questions about linear equivalence of divisors on algebraic curves. The latter has application to Computer Algebra, notably to the determination of antiderivatives of algebraic functions (Trager (1984), Bronstein (1990)), or the characterization of algebraic solutions of linear differential equations (Baldassarri and Dwork (1979), Compoint and Singer (1998)). In the context of Abel's map, one is led to compute integrals of algebraic functions along paths ending at critical points. Series expansions above critical points are definitely useful for this task (see Deconinck and Patterson (2007)). Moreover, representing algebraic functions as truncated Puiseux series could lead to error bounds for the integrals.

Unfortunately, applying a floating point Newton-Puiseux algorithm (see Section 3) to compute Puiseux series above a critical point is doomed to failure. Indeed, if the critical point $x_0$ is replaced with an approximation, expansion algorithms return approximate series with very small convergence discs and do not retain important information, such as ramification indices. Therefore, the output is not aidful.

On the other hand, coefficient growth considerably slows down symbolic methods. Since the degree of $\Delta_F$ is in $O(D^2)$, a critical point $x_0$ may be an algebraic number with large degree. Puiseux series coefficients above $x_0$ belong to a finite extension of $K$ whose degree over $K$ may be in $O(D^3)$. For $D = 10$, the extension may already be excessively large. Moreover, when these coefficients are expressed as linear combinations over $\mathbb{Q}$, the size of the rational numbers involved may also be overwhelming. Floating point evaluation of such coefficients must, in some cases, be performed with a high number of digits because spectacular numerical cancellations occurs (see examples in Poteaux (2007)). For instance, the polynomial of degree 6 in $Y$ $F(X, Y) = (Y^3 - X)((Y - 1)^2 - X)(Y - 2 - X^2) + X^2 Y^5$ has a discriminant $\Delta_F(X) = X^3 P(X)$, where $P(X)$ is an irreducible polynomial of degree 23 over $\mathbb{Q}$. Rational Puiseux series (see Section 2) above $P(X)$ have coefficients in a degree 23 extension of $\mathbb{Q}$. Rational numbers with 136 digits appear in the first term of the expansions. Walsh has shown that, for any $\epsilon > 0$, the singular part of Puiseux series can be computed using $O(d^{32+\epsilon} n^{4+\epsilon} \log h^{2+\epsilon})$ bit operations (Walsh (2000)), where $h$ is the height of $F$. Although this bound is probably not sharp, it is not encouraging and tends to confirm experimental observations.

To alleviate these problems, we introduce a symbolic-numeric approach : exact important information is first obtained by means of computation modulo a well chosen prime number $p$, then this information is used to guide floating point computation. The coefficient size is therefore kept under control while numerical instability is reduced. Exact important data, such as ramification indices and intersection multiplicities of branches, are preserved. Experimental evidences reported in (Poteaux (2007)) seem to validate this approach.

This paper presents several contributions :
- We introduce "generic Newton polygons" and "polygon trees" (Section 3). The latter capture precisely the symbolic information needed for floating point computations.
- We study modular reduction of Puiseux series and rational Puiseux expansions. This leads to a fully proved and easy to check criteria for the choice of a "good prime" $p$ such that polygon trees can be obtained using modular arithmetic (Section 4). We rely on technical results that are proven or recalled in Section 2 and 3.
- Complexity estimates for the computation of singular parts of Puiseux series over finite fields are given in Section 5. They improve previously published results.
- Finally, we study the bit-complexity of a Monte Carlo approach for the symbolic part of our symbolic-numeric method in Section 6.

Obtaining floating point Puiseux series from polygon trees is not a trivial task. The principles of our technique are briefly described in (Poteaux (2007)). Several variants, as well as round-off error control and overall complexity, are still under investigation. These questions are left to a forthcoming paper. Moreover, an implementation improving the prototype of (Poteaux (2007)) is being developed.

Modular methods are extensively used in Computer Algebra to avoid intermediate coefficient swell (via Hensel lifting or the Chinese Remainder Theorem, see for instance von zur Gathen and Gerhard (1999)). However, a "reduce mod $p$ and lift" or "reduce mod $p_i$ and combine" method would not help much in this case since we are not facing an intermediate coefficient growth problem, but an intrinsically large symbolic output. Modular methods are much less common when it comes to directly obtaining numerical results. We are aware of an attempt by Jean-Charles Faugère to control Gröbner bases computations that does not seem to have been successful so far (personal communication). We therefore claim some originality with this approach.

To conclude this introduction, we introduce notations and assumptions that will be used throughout the paper. We also recall well-known facts :
- All fields considered in the paper are commutative.
- If $L$ is a field, $\overline{L}$ will denote an algebraic closure of $L$.
- For each positive integer $e$, $\zeta_e$ is a primitive $e$-th root of unity in $\overline{L}$. Primitive roots are chosen so that $\zeta_{ab}^b = \zeta_a$.

- $v_X$ denotes the $X$-adic valuation of the fractional power series field $L((X^{1/e}))$, normalized with $v_X(X) = 1$. If $S \in L((X^{1/e}))$, we denote by $\text{tc}(S)$ the trailing coefficient of $S$, namely $S = \text{tc}(S) X^{v_X(S)} + $ terms of higher order.
- If $S = \sum_k \alpha_k X^{k/e}$ is a fractional power series in $L((X^{1/e}))$ and $r$ is a rational number, $\widetilde{S}^r$ denotes the truncated power series $\widetilde{S}^r = \sum_k^N \alpha_k X^{k/e}$ where $N = \max\{k \in \mathbb{N} \mid \frac{k}{e} \leq r\}$. We generalize this notation to elements of $L((X^{1/e}))[Y]$ by applying it coefficient-wise. In particular, if $H \in L[[X]][Y]$ is defined as $H = \sum_i (\sum_{k \geq 0} \alpha_{ik} X^k) Y^i$, then $\widetilde{H}^r = \sum_i (\sum_{k=0}^{\lfloor r \rfloor} \alpha_{ik} X^k) Y^i$.
- The discriminant of a univariate polynomial $U$ is denoted by $\Delta_U$. If $U$ is a multivariate polynomial, the context will always allow to identify the variable.
- If $U(T)$ (resp. $V(T)$) is a monic separable univariate polynomials of degree $s$ (resp. $r$) with roots $\{u_1, \ldots, u_s\}$ (resp. $\{v_1 \ldots, v_r\}$), then :

$$\Delta_U = \prod_{\substack{1 \leq i,j \leq s \\ i \neq j}} (u_i - u_j) \qquad \text{Resultant}(U, V) = \prod_{\substack{1 \leq i \leq s \\ 1 \leq j \leq r}} (u_i - v_j). \tag{1}$$

- If $U$ is a monic univariate polynomial which admits a factorization into a product of monic polynomials $U = \prod_{i=1}^r U_i$, then :

$$\Delta_U = \prod_{i=1}^r \Delta_{U_i} \prod_{\substack{1 \leq i,j \leq r \\ i \neq j}} \text{Resultant}(U_i, U_j). \tag{2}$$

- For each $i$ with $0 \leq i \leq e-1$, we denote $[i, e]$ the automorphism :

$$[i, e] : \overline{L}((X^{1/e})) \to \overline{L}((X^{1/e}))$$
$$X^{1/e} \mapsto \zeta_e^i X^{1/e}$$

When the ramification index can be deduced from the context, we shall simply write $[i]$ instead of $[i, e]$. If $S \in \overline{L}((X^{1/e})$, the image of $S$ under $[i]$ is denoted by $S^{[i]}$. This notation extends naturally to any polynomial with coefficient in $\overline{L}((X^{1/e})$. It is obvious that elements of the subfield $\overline{L}((X))$ are invariant under $[i]$.
- Let $f$ be a polynomial in $L[T]$ with squarefree factorization $f = \prod_{i=1}^r f_i^{k_i}$ (that is, the $k_i$ are pairwise distinct positive integers and the $f_i$ are positive degree polynomials with $\gcd(f_i, f_j) = 1$ if $i \neq j$). We associate to $f$ the partition of $\deg f$ denoted $[f] = (k_1^{\deg f_1} \ldots k_r^{\deg f_r})$. Namely, the multiplicity $k_i$ is repeated $\deg f_i$ times in the decomposition of $\deg f$.
- If $H \in L[X, Y]$, then $H_X$ and $H_Y$ are the formal partial derivatives of $H$.
- For a multivariate polynomial $H(\underline{X}) = \sum_{\underline{k}} \alpha_{\underline{k}} \underline{X}^{\underline{k}} \in \mathbb{C}[\underline{X}] = \mathbb{C}[X_1, \ldots, X_n]$, where $\underline{k}$ is a multi-index, we denote :

$$||H||_\infty = \max_{\underline{k}} \{|\alpha_{\underline{k}}|\} \qquad ||H||_2 = \sqrt{\sum_{\underline{k}} \{|\alpha_{\underline{k}}|^2\}}.$$

## 2. Puiseux series

We need to state results over more general fields than $K$. Throughout the section, $L$ stands for a field of characteristic $p \geq 0$ and $F$ belongs to $L[X, Y]$. Otherwise, we keep the assumptions and notations of Section 1. We also impose the condition :

$$p = 0 \quad \text{or} \quad p > d = \deg_Y(F) \tag{3}$$

Up to a change of variable $X \leftarrow X + x_0$, we assume that the critical point is $X = 0$.

*2.1. Classical Puiseux series*

In this part, we review classical results about Puiseux series. We begin with :

**Theorem 1** (Puiseux). *Let $H$ be squarefree polynomial of $L[X,Y]$ such that $\deg_Y(H) = d > 0$.*
- *If condition (3) is satisfied, there exist positive integers $e_1, \ldots, e_s$ satisfying $\sum_{i=1}^{s} e_i = d$ such that $H$ (viewed as a polynomial in $Y$) has $d$ distinct roots in $\overline{L((X))}$ which can be written :*

$$S_{ij}(X) = \sum_{k > -\infty} \alpha_{ik}\, \zeta_{e_i}^{jk}\, X^{\frac{k}{e_i}}$$

*for $1 \leq i \leq s$ and $0 \leq j \leq e_i - 1$. Moreover, the set of coefficients $\{\alpha_{ik}\}$ is included in a finite algebraic extension of $L$.*
- *If $p = 0$, then :*

$$\overline{L(X)} \subset \overline{L((X))} = \bigcup_{e \in \mathbb{N}^*} \overline{L}((X^{1/e}))$$

**Definition 2.** These $d$ fractional Laurent series are called *Puiseux series of $H$ above $0$*. The integer $e_i$ is the *ramification index* of $S_{ij}$. If $e_i > 1$, then $S_{ij}$ is *ramified*. If $S_{ij} \in \overline{L}[[X^{1/e_i}]]$, we say that $S_{ij}$ is *defined at $X = 0$*. If $S_{ij}(0) = 0$, we say that $S_{ij}$ *vanishes at $X = 0$*.

An arbitrary number of terms of all Puiseux series can be effectively computed using Newton-Puiseux algorithm (see Section 3).

For each positive integer $e \leq d$, our hypothesis (3) about the characteristic $p$ of $L$ imply that the Galois group $\mathbb{G}_e$ of $\overline{L}((X^{1/e}))/\overline{L}((X))$ is cyclic and generated by [1] : $X^{1/e} \mapsto \zeta_e X^{1/e}$. Hence, $\mathbb{G}_{e_i}$ permutes cyclically the elements of the set $S_i = \{S_{ij}(X)\}_{0 \leq j \leq e_i - 1}$.

**Definition 3.** We call $S_i$ *a cycle of $H$ above $0$*. If an element of $S_i$ (thus, all elements) vanishes at $X = 0$, we say that *the cycle vanishes at $X = 0$*.

Since the $S_{ij}$ ($0 \leq j \leq e_i - 1$) can be quickly recovered (both symbolically and numerically) from any element of $S_i$, it is sufficient for our purposes to compute a set of representatives for the cycles of $H$.

**Definition 4.** The *regularity index* $r_{ij}$ of $S_{ij}$ in $H$ is the least integer $N$ such that $\widetilde{S_{ij}}^{\frac{N}{e_i}} = \widetilde{S_{uv}}^{\frac{N}{e_i}}$ implies $(u,v) = (i,j)$. The truncated series $\widetilde{S_{ij}}^{\frac{r_{ij}}{e_i}}$ is called the *singular part of $S_{ij}$ in $H$*.

In other words, $r_{ij}$ is the smallest number of terms necessary to distinguish $S_{ij}$ from the other Puiseux series above $0$. It is worth noting that $r_{ij}$ depends not only on $S_{ij}$, but also on $H$ since $H$ is not assumed irreducible in $L[X,Y]$. See examples in Section 3.2.

If the singular part of a Puiseux series is known, a change of variable yields a bivariate polynomial for which remaining terms of the series can be computed "fast" using quadratic Newton iterations (Kung and Traub (1978), von zur Gathen and Gerhard (1999)). Newton iterations can be applied to series with floating point coefficients, therefore we focus on the computation of the singular parts of the $S_{ij}$. Since it can be shown that all elements of a cycle $S_i$ have the same regularity index, that we denote $r_i$, the problem reduces to the determination of the singular part of a representative of $S_i$ for $1 \leq i \leq s$.

Since $F$ is monic, all $S_{ij}$ are defined at $X = 0$. In other words, they are fractional power series. In this case, when $L \subset \mathbb{C}$, the $S_{ij}$ have a convergence radius which is at least equal to the distance from 0 to the nearest (nonzero) critical point. If we choose a determination for the $e_i$-th root functions, the $S_{ij}$ define $d$ analytic functions in any domain $\mathcal{D}$ that is included in the convergence disc and does not intersect the branch cut. To evaluate accurately these functions in $\mathcal{D}$, we need to :

- Control truncation orders of Puiseux series. Bounds are given in (Poteaux (2007)).
- Compute efficiently floating point approximation of the truncated $S_{ij}$. The principles and a preliminary implementation are presented in (Poteaux (2007)). The present paper details the symbolic part of our method.
- Give error bounds for the approximations of the $\alpha_{ik}$ and study the algorithm numerical stability. This topic has not been addressed yet.

### 2.2. The characteristic of a Puiseux series

We derive relations between the discriminant of $F$ and particular coefficients of its Puiseux series that we shall use to define a "good reduction" criterion. Let

$$S(X) = \sum_{i=0}^{\infty} \alpha_i X^{i/e}$$

denote a Puiseux series of $F$ with ramification index $e > 1$. We define a finite sequence $(B_0, R_0), (B_1, R_1), \ldots, (B_g, R_g))$ of integer pairs as follows :
- $R_0 = e$, $B_0 = 0$.
- If $R_{j-1} > 1$, we set $B_j = \min \{i > B_{j-1} \mid \alpha_i \neq 0 \text{ and } i \not\equiv 0 \pmod{R_{j-1}}\}$, $R_j = \gcd(B_j, R_{j-1})$. If $R_{j-1} = 1$, we stop and set $g = j - 1$. Note that $g \geq 1$ and $R_g = 1$.

Finally, we set $Q_j = R_{j-1}/R_j$, $M_j = B_j/R_j$ for $(1 \leq j \leq g)$ and define $H_j$ to be the largest non negative integer so that $B_j + H_j R_j < B_{j+1}$ for $0 \leq j \leq g-1$. It is clear that $e = Q_1 Q_2 \cdots Q_g$ and $M_j$ is an integer prime to $Q_j$.

With these notations, and up to a new indexing of the coefficients, $S$ can be written in the form :

$$\begin{aligned}
S(X) = {} & \sum_{j=0}^{H_0} \alpha_{0,j} X^j \\
& + \gamma_1 X^{\frac{M_1}{Q_1}} && + \sum_{j=1}^{H_1} \alpha_{1,j} X^{\frac{M_1+j}{Q_1}} \\
& + \gamma_2 X^{\frac{M_2}{Q_1 Q_2}} && + \sum_{j=1}^{H_2} \alpha_{2,j} X^{\frac{M_2+j}{Q_1 Q_2}} \\
& + \cdots && + \cdots \\
& + \gamma_g X^{\frac{M_g}{Q_1 Q_2 \cdots Q_g}} && + \sum_{j=1}^{\infty} \alpha_{g,j} X^{\frac{M_g+j}{Q_1 Q_2 \cdots Q_g}}
\end{aligned} \qquad (4)$$

In the expression above, the monomials of $S$ are ordered by increasing (rational) degree.

**Definition 5** ((see Zariski (1981) or Brieskorn and Knörrer (1986))). The *characteristic* of $S$ is the tuple of integers $(e; B_1, \ldots, B_g)$. The *characteristic coefficients* are the elements of the sequence $(\gamma_1, \ldots, \gamma_g)$ and *characteristic monomials* are the corresponding monomials of $S$.

**Proposition 6.** *Assume that hypothesis (3) is satisfied. Let $G(X,Y)$ be the minimal polynomial over $\overline{L}((X))$ of a ramified Puiseux series $S \in \overline{L}[[X^{1/e}]]$ as above. Let $\Delta_G(X)$ be the discriminant of $G$ with respect to $Y$. Then :*

$$\text{tc}(\Delta_G) = \pm (\prod_{i=1}^{g} Q_i^{R_i} \prod_{i=1}^{g} \gamma_i^{R_{i-1}-R_i})^e \tag{5}$$

$$v_X(\Delta_G) = \sum_{i=1}^{g} B_i(R_{i-1} - R_i) \tag{6}$$

**Proof.** We first introduce the notation $v = v_X(\Delta_G)$ and $\theta = \text{tc}(\Delta_G)$. The conjugates of $S$ over $\overline{L}((X))$ are $\{S^{[i]}\}_{0 \leq i \leq e-1}$, therefore :

$$\Delta_G = \prod_{\substack{0 \leq i,j \leq e-1 \\ i \neq j}} (S^{[i]} - S^{[j]}).$$

From this relation and (5), we note that $v$ depends only on the contribution of the terms $X^{B_i/e} = X^{M_i/(Q_1 \cdots Q_i)}$. Hence, $v$ is determined by the exponent of $\gamma_i$ in $\theta$. Therefore, if (5) is true, so is (6) since :

$$v = \sum_{i=1}^{g} e\,(R_{i-1} - R_i) \frac{B_i}{e} = \sum_{i=1}^{g} B_i(R_{i-1} - R_i).$$

In order to prove (5), we proceed by induction on $g$. For each positive integer $r$ let $\delta_r$ be the discriminant of $X^r - 1$, that is $\delta_r = \pm r^r$.

If $g = 1$, the expansion of $\Delta_G$ in increasing fractional power of $X$ is :

$$\Delta_G = \prod_{\substack{0 \leq i,j \leq e-1 \\ i \neq j}} \left( \gamma_g(\zeta_e^{M_g i} - \zeta_e^{M_g j}) X^{M_g/Q_g} + \cdots \right)$$

$$= \gamma_g^{e(e-1)} \left( \prod_{\substack{0 \leq i,j \leq e-1 \\ i \neq j}} (\zeta_e^{M_g i} - \zeta_e^{M_g j}) \right) X^{e(e-1)M_g/Q_g} + \cdots$$

Since $M_g$ is prime to $Q_g$ and $Q_g = e$, $\zeta_e^{M_g}$ is a primitive $e$-th root of unity. We obtain $\theta = \delta_e \gamma_g^{e(e-1)} = \pm Q_g^{Q_g} \gamma_g^{e(R_{g-1}-R_g)}$ as expected.

We now assume that $g > 1$. To simplify notations, we set $Q = Q_1$ and $R = R_1 = Q_2 \cdots Q_g$. We define $H \in \overline{L}[[X^{1/Q}]][Y]$ as follows :

$$H = \prod_{i=0}^{R-1} (Y - S^{[iQ]}).$$

Since $[Q] = [Q, e]$ generates the Galois group of $\overline{L}((X^{1/e}))$ over $\overline{L}((X^{1/Q}))$, $H$ is the minimal polynomial of $S$ over $\overline{L}((X^{1/Q}))$. Moreover, the factorization of $G$ over $\overline{L}((X^{1/Q}))$ is given by :

$$G = \prod_{i=0}^{Q-1} H^{[i]}.$$

Using relation (2), we obtain $\Delta_G = \Pi_1 \Pi_2$ where :

$$\Pi_1 = \prod_{i=0}^{Q-1} \Delta_{H^{[i]}} \qquad \Pi_2 = \prod_{\substack{0 \leq i,j \leq Q-1 \\ i \neq j}} \text{Resultant}(H^{[i]}, H^{[j]}).$$

We need to evaluate the contribution to $\theta$ of $\Pi_1$ and $\Pi_2$. We first consider $\Pi_1$. Let $U(X, Y) = H(X^Q, Y)$ be the minimal polynomial of $S(X^Q)$ over $\overline{L}((X))$. Since $U$ has characteristic $(R; B_2, \ldots, B_g)$, our induction hypothesis yields :

$$\Delta_U = (\pm \prod_{i=2}^{g} Q_i^{R_i} \prod_{i=2}^{g} \gamma_i^{R_{i-1} - R_i})^R X^u + \cdots$$

for some positive integer $u$. Therefore :

$$\Delta_{H^{[j]}} = \zeta_e^{uRj} (\pm \prod_{i=2}^{g} Q_i^{R_i} \prod_{i=2}^{g} \gamma_i^{R_{i-1} - R_i})^R X^{\frac{u}{Q}} + \cdots$$

Since $QR = e$ and $\zeta_e^{uRj} = \zeta_Q^{uj}$ the contribution of $\Pi_1$ to $\theta$ is :

$$\pm (\prod_{i=2}^{g} Q_i^{R_i} \prod_{i=2}^{g} \gamma_i^{R_{i-1} - R_i})^e \tag{7}$$

We now estimate the contribution of $\text{Resultant}(H^{[i]}, H^{[j]})$. Each difference of roots in the product defining the resultant has the form :

$$\gamma_1 (\zeta_Q^{M_1 i} - \zeta_Q^{M_1 j})(X^{M_1/Q_1} + \ldots)$$

and there are $R^2$ such differences. Since there are $Q(Q-1)$ resultants in the product and $\zeta_Q^{M_1}$ is a primitive $Q$-th root of unity, we conclude that the contribution of $\Pi_2$ to $\theta$ is :

$$\gamma_1^{R^2 Q(Q-1)} (\prod_{\substack{0 \leq i,j \leq Q-1 \\ i \neq j}} \zeta_Q^{M_1 i} - \zeta_Q^{M_1 j})^{R^2} = \gamma_1^{R^2 Q(Q-1)} \delta_Q^{R^2} = \pm Q_1^{eR_1} \gamma_1^{e(R_0 - R_1)}$$

Combining the last expression with (7) gives (5). □

**Remark 7.** The value of $v_X(\Delta_F)$ is well-known (see for instance (Zariski (1981))). It can be expressed as the sum of a differential exponent and of a conductor degree. In Singularity Theory, it also has an interpretation in terms of "infinitely near point" multiplicities. However, we have not found in the literature an expression for $\text{tc}(\Delta_F)$.

*2.3. Rational Puiseux expansions*

In order to perform computations in the smallest possible extension of $L$ and to take advantage of conjugacy over $L$, Duval introduced the notion of "rational Puiseux expansions over $L$" (Duval (1987)). This arithmetical concept is irrelevant in the context of floating point computations, but will prove useful for expansions over finite fields.

**Remark 8.** A different definition of "rational Puiseux expansions over $L$" appeared in (Duval (1989)) and (Walsh (1999)). The definition given therein corresponds to "rational Puiseux expansions over $\overline{L}$" in the sense of (Duval (1987)) and in the sense of this paper.

**Definition 9.** Let $H$ be a polynomial in $L[X,Y]$ with $\deg_Y H > 0$. A *parametrization* $R(T)$ of $H$ is a pair of non constant power series $R(T) = (X(T), Y(T)) \in \overline{L}((T))^2$ such that $H(X(T), Y(T)) = 0$ in $\overline{L}((T))$. The parametrization is *irreducible* if there is no integer $u > 1$ such that $R(T) \in \overline{L}((T^u))^2$. The *coefficient field* of $R(T)$ is the extension of $L$ generated by the coefficients of $X(T)$ and $Y(T)$.

Assume for a moment that $H$ is irreducible in $L[X,Y]$ so that $\mathcal{K} = L(X)[Y]/(H)$ is an algebraic function field. A parametrization $R(T) = (X(T), Y(T))$ induces a field morphism :

$$\phi_R : \quad \mathcal{K} \to \overline{L}((T))$$
$$f(X,Y) \mapsto f(X(T), Y(T))$$

Composing $\phi_R$ with the valuation $v_T$ of $\overline{L}((T))$, we obtain a valuation of $\mathcal{K}$ that we denote again by $v_T$. It is easily seen that the set $\mathfrak{P}_R = \{f \in \mathcal{K} \,|\, v_T(f) > 0\}$ is a *place* of $\mathcal{K}$ in the sense of (Chevalley (1951)) and that $V_R = \{f \in \mathcal{K} \,|\, v(f) \geq 0\}$ is the corresponding V-ring of $\mathcal{K}$. We recall that $\mathfrak{P}_R$ is the unique maximal ideal of $V_P$ and that the *residue field* of $\mathfrak{P}_R$ is $V_R/\mathfrak{P}_R$, which can be viewed as a finite algebraic extension of $L$. Therefore, we obtain a mapping $\Psi$ from the set of parametrizations of $F$ onto the set of places of $\mathcal{K}$. Reciprocally, to each place $\mathfrak{P}$ of $\mathcal{K}$ corresponds parametrizations of $H$.

Let us denote by $\{\mathfrak{P}_i\}_{1 \leq i \leq r}$ the places of $\mathcal{K}$ dividing $X$ and by $k_i$ the residue field of $\mathfrak{P}_i$.

**Definition 10** (Rational Puiseux expansions)**.**
- Assume that $H$ is irreducible in $L[X,Y]$. A *system of L-rational Puiseux expansions above 0 of $H$* is a set of irreducible parametrizations $\{R_i\}_{1 \leq i \leq r}$ of the form :

$$R_i(T) = (X_i(T), Y_i(T)) = \left(\gamma_i T^{e_i}, \sum_{k > -\infty} \beta_{ik} T^k\right) \in \overline{L}((T))^2$$

  with $e_i > 0$ such that :
  (i) $\Psi$ is one-to-one from $\{R_i\}_{1 \leq i \leq r}$ to $\{\mathfrak{P}_i\}_{1 \leq i \leq r}$. We assume that the $\mathfrak{P}_i$ are numbered so that $\mathfrak{P}_i = \mathfrak{P}_{R_i} = \Psi(R_i)$.
  (ii) The coefficient field of $R_i$ is isomorphic to $k_i$.
- Assume that $H$ is squarefree. A *system of L-rational Puiseux expansions above 0 of $H$* is the union of systems of $L$-rational Puiseux expansions for the irreducible factors of $H$ in $L[X,Y]$.

**Definition 11.** We say that $R_i$ is *defined at $T = 0$* if $Y_i \in \overline{L}[[T]]$. In this case, the *center* of $R_i$ is the pair $(X_i(0), Y_i(0)) \in \overline{L}^2$.

The classical formula relating degrees of residue fields and ramification indices of an algebraic function field (see Chevalley (1951)) translates into :

**Theorem 12.** *Let $H \in L[X,Y]$ be squarefree and $\deg_Y H = d > 0$. Let $\{R_i\}_{1 \leq i \leq r}$ be a system of L-rational Puiseux expansion above 0 for $H$. Let $f_i$ stand for the degree over $L$ of the coefficient field of $R_i$. Then :*

$$\sum_{i=1}^{r} e_i f_i = d.$$

Classical Puiseux series can be readily deduced from a system of rational Puiseux expansions :

(1) $R_i$ has exactly $f_i$ conjugates over $L$, that we denote $R_i^\sigma$ ($1 \leq \sigma \leq f_i$).

$$R_i^\sigma(T) = (X_i^\sigma(T), Y_i^\sigma(T)) = \left(\gamma_i^\sigma T^{e_i}, \sum_{k > -\infty} \beta_{ik}^\sigma T^k\right)$$

(2) Each $R_i^\sigma$ yields a Puiseux series $Y_i^\sigma((X/\gamma_i^\sigma)^{1/e_i})$. The set of all such series form a set of representatives for set of cycles $\{S_l\}_{1 \leq l \leq s}$ of $H$ above 0.
(3) The $d$ Puiseux series are finally obtained using the action of $\mathbb{G}_{e_i}$, $1 \leq i \leq s$.

In particular, classical Puiseux series that are defined at $X = 0$ (resp. that vanish at $X = 0$) correspond to rational Puiseux expansions defined at $T = 0$ (resp. centered at $(0,0)$).

Again, we note that regularity indices for all Puiseux series corresponding to the same rational Puiseux expansion are equal. Therefore, we define the singular part of a rational Puiseux expansion $R_i$ to be the pair :

$$\left(\gamma_i T^{e_i}, \sum_{k > -\infty}^{r_i} \beta_{ik} T^k\right)$$

where $r_i$ is the regularity index of a Puiseux series associated to $R_i$.

Finally, we note that since $F$ is monic, rational Puiseux expansions of $F$ above zero are all defined at $T = 0$, that is $R_i(T) \in \overline{L}[[T]]^2$.

## 3. Symbolic algorithms for Puiseux series

In this section, we describe two algorithms to compute singular parts of Puiseux series and rational Puiseux expansions. Both methods are used by our symbolic-numeric approach. We also explain how coefficients computed by the two methods are related and conclude by comments about coefficient growth.

Throughout the section, $L$ stands again for a field of characteristic $p \geq 0$ and $F$ is a polynomial $L[X,Y]$ such that condition (3) is satisfied. Moreover, we keep the assumptions and notations of Section 1.

Newton polygons and characteristic polynomials are the crucial tools. We first recall well-known definitions and introduce a variant that will prove more convenient and powerful.

## 3.1. Generic Newton polygons and characteristic polynomials

Assume that $H(X,Y) = \sum_{i,j} a_{ij} X^j Y^i$ is a polynomial of $L[[X]][Y]$ such that $H(0,Y) \neq 0$. The Newton polygon of $H$ is classically defined as follow :

**Definition 13.** Denote by $\mathcal{I}(H)$ the nonnegative integer $v_Y(H(0,Y))$ and by $\mathcal{H}$ the convex hull of $\mathrm{Supp}(H) = \{(i,j) \in \mathbb{N}^2 \,|\, a_{ij} \neq 0\}$. The *Newton Polygon* $\mathcal{N}(H)$ of $H$ is the lower part of $\mathcal{H}$. Namely :
- If $H(X,0) \neq 0$, $\mathcal{N}(H)$ is formed by the sequence of edges of $\mathcal{H}$ closest to the origin and joining $(0, v_X(H(X,0)))$ to $(\mathcal{I}(H), 0)$,
- If $H(X,0) = 0$, $(0, v_X(H(X,0)))$ is replaced by the leftmost point of $\mathcal{H}$ with smallest $j$-coordinate.

In particular, a vertical or horizontal edge is not considered part of $\mathcal{N}(H)$. The Newton polygon may be reduced to a single point. For instance $H(X,Y) = Y$ yields the trivial polygon $(1,0)$.

We now introduce a slightly different object, that we call *generic Newton polygon* for reasons explained later. This variation allows a homogeneous treatment of finite series, clearer specifications for the algorithms and simplifies wording and proofs of results regarding modular reduction.

**Definition 14.** The *generic Newton polygon* $\mathcal{GN}(H)$ is obtained by restricting $\mathcal{N}(H)$ to edges with slope no less than $-1$ and by joining the leftmost remaining point to the vertical axis with an edge of slope $-1$.

In other words, we add a fictitious point $(0, j_0)$ to $\mathrm{Supp}(H)$ so as to mask edges with slope less than -1.

**Example 15.** Consider $H_1(X,Y) = Y^7 + X^2 Y^2 + XY^4 + X^8 + X^6 + Y^3 X^2 + Y^5 X + Y^3 X^4$. In Figure 3.1, the support of $H_1$ is represented by crosses, $\mathcal{GN}(H_1)$ is drawn with plain lines while the masked edge of $\mathcal{N}(H_1)$ is represented by a dotted line.

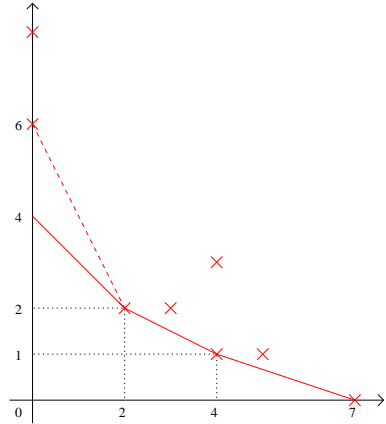

Fig. 1. $\mathcal{GN}(H_1)$ versus $\mathcal{N}(H_1)$

**Example 16.** Consider $H_2(X,Y) = Y^8 + 3X^2Y^3 + XY^5 + YX^8 + 2X^6 + 4Y^2X^3 + Y^5X^2 + Y^5X^3 + Y^3X^4$ and Figure 3.1. The edge with slope -1 is prolongated until the vertical axis.

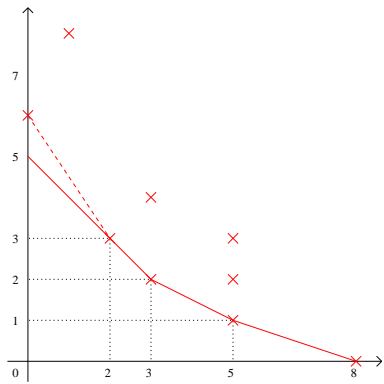

Fig. 2. $\mathcal{GN}(H_2)$ versus $\mathcal{N}(H_2)$

**Example 17.** Assume that $H_3(X,Y) = Y$. Then $\mathcal{GN}(H_3)$ is formed of a unique edge joining $(0,1)$ to $(1,0)$.

The first stage of the algorithms requires a special treatment. To this effect, it is convenient to introduce the following definition :

**Definition 18.** The exceptional Newton polygon $\mathcal{EN}(H)$ is the unique horizontal edge $[(0,0),(\deg_Y(H(0,Y)),0)]$.

In particular, $\mathcal{EN}(F) = [(0,0),(d,0)]$ since $F$ is monic.

To an edge $\Delta$ of $\mathcal{GN}(H)$ (or $\mathcal{N}(H)$, $\mathcal{EN}(H)$) corresponds three nonnegative integers $q$, $m$ and $l$ with $q$ and $m$ coprime such that $\Delta$ is on the line $qj + mi = l$. If $\Delta$ is the horizontal edge of $\mathcal{EN}(H)$, $m = l = 0$ and we choose $q = 1$.

**Definition 19.** We define the *characteristic polynomial* $\phi_\Delta$ :

$$\phi_\Delta(T) = \sum_{(i,j)\in\Delta} a_{ij} T^{\frac{i-i_0}{q}}$$

where $i_0$ is the smallest value of $i$ such that $(i,j)$ belongs to $\Delta$.

Note that if $\mathcal{N}(H)$ is used, $\phi_\Delta(T)$ cannot vanish at $T = 0$, while $\mathcal{GN}(H)$ allows such cancellation if $\Delta$ is a fictitious edge (or contain a fictitious part). In this case, the multiplicity of 0 as a root of $\phi_\Delta(T)$ is the length of the fictitious edge (or portion of edge) added. For $\mathcal{EN}(H)$, 0 can also be a root of the characteristic polynomial.

The next two lemmas recall the relation between Newton polygons of $H$ and Newton polygons of its factors in $L[[X]][Y]$ :

**Lemma 20.** *If $H$ is an irreducible polynomial of $L[[X]][Y]$ and $H(0,0) = 0$, then $\mathcal{GN}(H)$ has a unique edge $\Delta$ and $\phi_\Delta$ has a unique root.*

**Proof.** This is well-known for classical Newton polygons (Brieskorn and Knörrer (1986)). The extension to generic polygons is straightforward. □

**Lemma 21.** *Let $H_1$ and $H_2$ be elements of $L[[X]][Y]$. Then, $\mathcal{GN}(H_1 H_2)$ results from joining together the different edges of $\mathcal{GN}(H_1)$ and $\mathcal{GN}(H_2)$, suitably displaced. Moreover, the characteristic polynomial of an edge $\Delta$ with slope $-m/q$ of $\mathcal{GN}(H_1 H_2)$ is the product of the characteristic polynomials associated with edges of $\mathcal{GN}(H_1)$ and $\mathcal{GN}(H_2)$ of slope $-m/q$. In particular, if $H_1(0,0) \neq 0$, so that $\mathcal{GN}(H_1)$ is reduced to the point $(0,0)$, then $\mathcal{GN}(H_1 H_2) = \mathcal{GN}(H_2)$.*

**Proof.** For classical Newton polygons, see (Brieskorn and Knörrer (1986)). For generic Newton polygons, proceed as follow : If necessary, add a monomial $cX^{n_1}$ (resp. $cX^{n_2}$) to $H_1$, where $c$ is an indeterminate, so that $\mathcal{GN}(H_i) = \mathcal{N}(H_i)$. Then, apply the result for the classical case and set $c = 0$ to recover $\mathcal{GN}(H_1 H_2)$. □

Both algorithms below perform successive changes of variable, determined by $(q,m,l)$ and the roots of $\phi_\Delta$. They return a set of triplets $\{(G_i(X,Y), P_i(X), Q_i(X,Y))\}_i$ such that :
- $G_i, P_i, Q_i \in \overline{L}[X,Y]$,
- $P_i(X)$ is a monomial of the form $\lambda_i X^{e_i}$,
- $Q_i(X,Y) = Q_{i0}(X) + Y X^{r_i}$, where $r_i$ is the regularity index of the expansion and $(P_i(T), Q_{i0}(T))$ is the singular part of a parametrization of $F$,
- There exist nonnegative integers $L_i$ such that $G_i(X,Y) = F(P_i(X), Q_i(X,Y))/X^{L_i}$, $G_i(0,0) = 0$ and $G_{iY}(0,0) \neq 0$.

By the formal Implicit Function Theorem, the latter conditions ensure that there exists a unique power series $S$ such that $G_i(X, S(X)) = 0$ and $S(0) = 0$. The corresponding parametrization of $F$ is therefore $R_i(T) = (P_i(T), Q_i(T, S(T)))$. The power series $S$ can be computed using "fast" techniques (Kung and Traub (1978)). It may also happen that $Y$ divides $G_i$, in which case the expansion is finite. Therefore, we can consider that such a triplet represents a Puiseux series or a rational Puiseux expansion.

*3.2. Classical Puiseux series*

We first give a variant of the classical Newton-Puiseux algorithm (see Walker (1978)) to compute the singular parts of Puiseux series. Our floating point algorithm is based on this pattern (see Poteaux (2007)). This version returns exactly one representative for each cycle above 0. The algorithm is presented in a recursive setting and the output is given in a parametric form by mean of triplets as above. The first (non recursive) call to the function must be treated differently since $\mathcal{EN}(H)$ must be used instead of $\mathcal{GN}(H)$, for reasons explained in Section 4.2. We assume that a mechanism is available to distinguish the initial call from recursive calls (adding a Boolean argument would work, for instance).

```
CNPuiseux(H)
Input:
     H : A squarefree polynomial of degree d ≥ 1 in L[X,Y], such that H(0,Y) ≠ 0.
Output:
```

  *A set of triplets $\{[G_i, P_i, Q_i]\}_i$, which form a set of representatives for :*
  *- all cycles of H above 0 defined at $X = 0$ for the initial call,*
  *- cycles of H that vanish at $X = 0$ for recursive calls.*

```
Begin
  If in a recursive call then
    N ← GN(H)
    If I(H) = 1 then Return {[H, X, Y]} End
  else
    N ← EN(H)
  End
  For each side Δ of N do
    Compute q, m, l and φ_Δ
    For each (distinct) root ξ of φ_Δ do
      α ← ξ^{1/q}
      H'(X, Y) ← H(X^q, X^m(α + Y))/X^l
      For each [G, P, Q] in CNPuiseux(H) do
        R ← R ∪ {[G, P^q, P^m(α + Q)]}
      End
    End
  End
  Return R
End.
```

In particular, when applied to the monic polynomial $F$, CNPuiseux returns a set of representatives for all the cycles of $F$ above 0.

**Example 22.** Set $F(X, Y) = Y^3 - X^5 \in \mathbb{Q}[X, Y]$. Applying CNPuiseux yields one triplet with $P_1(X) = X^3$ and $Q_1(X, Y) = X^0(0 + X^3(0 + X^2(1 + Y))) = X^5 + X^5Y$. The first null coefficient comes from the exceptional polygon $[(0, 0), (0, 3)]$. The second one correspond to the fictitious edge of $\mathcal{GN}(F)$ introduced at the first recursive call. This may seem inefficient, but these tricks have no impact on the complexity and clarifies arguments in Section 4. In practice, one may still use a classical Newton polygon if necessary.

**Example 23.** Consider $F(X, Y) = (Y - 1 - 2X - X^2)(Y - 1 - 2X - X^7) \in \mathbb{Q}[X, Y]$. We obtain two triplets with :

$$(P_1, Q_1) = (X, X^0(1 + X(2 + X(1 + Y)))) = (X, 1 + 2X + X^2 + X^2Y)$$
$$(P_2, Q_2) = (X, X^0(1 + X(2 + X(0 + Y)))) = (X, 1 + 2X + X^2Y)$$

Note that the generic Newton polygon allows to obtain immediately the regularity index of the series $X + X^7$ in $F$, which is 2. The classical polygon does not provide directly this information.

**Example 24.** Let $F$ be the product of the minimal polynomials over $\mathbb{Q}(X)$ of the series $X^{5/6} + X$ and $X^{5/6} + X^{11/12}$. We obtain two triplets with :

$$(P_1, Q_1) = (X^6, X^0(0 + X^5(1 + X(1 + Y)))) = (X^6, X^5 + X^6 + X^6Y)$$
$$(P_2, Q_2) = (X^{12}, X^0(0 + X^{10}(1 + X(1 + Y)))) = (X^{12}, X^{10} + X^{11} + X^{11}Y)$$

The regularity indices in $F$ are indeed 6 and 11.

The data $q, m, l$ come directly from the polygons. They need to be computed exactly, since for instance $q$ will contribute to the ramification index, which has to be obtained exactly. It is obvious that if $\xi$ is replaced by a numerical approximation, the change of variable in CNPuiseux will almost always produce a polynomial $H'$ with trivial Newton polygon, namely reduced to the unique point $(0,0)$. It will not be easy to recover the correct polygon, since we will have to decide which coefficients are approximations of 0 and should be ignored. Moreover :

**Proposition 25.** *Let $H$ be a polynomial satisfying the input hypotheses of* CNPuiseux.
- *The integer $\mathcal{I}(H)$ is the number of Puiseux series of $H$ above 0 vanishing at $X = 0$.*
- *The integer $\mathcal{I}(H')$ is equal to the multiplicity of $\xi$ in $\phi_\Delta$.*

**Proof.** See (Duval (1989)). □

The second assertion of Proposition 25 shows that $\phi_\Delta$ is not squarefree in general. In the presence of approximations, determining the distinct roots of $\phi_\Delta$ and their multiplicity may be difficult.

However, if we assume that all Newton polygons are obtained by some other means (which implies that multiplicities are also known) and provided as an input, then we can :

(1) Extract the approximate coefficients of $H$ which are meaningful to compute $\mathcal{GN}(H)$. The coefficients below $\mathcal{GN}(H)$ should be equal to 0 : discard them.
(2) Deduce an approximate $\phi_\Delta$.
(3) Find clusters of approximate roots of $\phi_\Delta$ with the expected multiplicities.
(4) For each cluster, deduce an approximate value of $\xi$, apply the numerical change of variable and proceed with the recursive call.

With this technique, we obtain approximate Puiseux series with correct ramification indices, that is correct combinatorial data. Our experiments show that approximations for the series coefficients are reasonably accurate, yielding accurate evaluations of algebraic functions in the neighborhood of critical points. To compute the exact data that are needed, we use computations modulo a well chosen prime number $p$ (see Section 4). A method to establish the correspondence between data modulo $p$ and numerical data is sketched in Poteaux (2007).

**Remark 26.** For any irreducible $F(X,Y) \in \overline{L}[[X]][Y]$, at each step of CNPuiseux, the polygon $\mathcal{N}$ has exactly one edge and the characteristic polynomial has a unique root. Moreover, the sequence of generic Newton polygons encountered depends only on the characteristic terms of the Puiseux series (see Section 2.2), and not on the other terms. In this sense, these polygons are truly "generic", since all $F$ with the same characteristic yield the same sequence of generic polygons.

## 3.3. Rational Puiseux expansions

We now describe an algorithm, due to Duval, to compute singular parts of rational Puiseux expansions above 0, that we shall use over finite fields. We need two auxiliary algorithms, for which we only provide specifications :

Factor($L$,$\phi$)

Input:
>   $L$ : a field
>   $\phi$ : a univariate polynomial in $L[T]$.

Output:
>   A set of pairs $\{(\phi_i, k_i)\}_i$ so that $\phi_i$ is irreducible in $L[T]$ and $\phi = \prod_i \phi_i^{k_i}$.

Bzout($q$,$m$)

Input:
>   $q, m$ : two positive integers

Output:
>   A pair of integers $(u, v)$ so that $uq - mv = 1$. If $q = 1$, enforce $v = 0$, and $u = 1$.

Algorithm RNPuiseux($L$,$H$)

Input:
>   $L$ : A field.
>   $H$ : A squarefree polynomial of degree $d \geq 1$ in $L[X, Y]$, such that $H(0, Y) \neq 0$.

Output:
>   A set of triplets $\{[G_i, P_i, Q_i]\}_i$, which form a set of representatives for :
>   - $L$-rational Puiseux expansions of $H$ above 0 defined at $T = 0$ for the initial call,
>   - $L$-rational Puiseux expansions of $H$ centered at $(0, 0)$ for recursive calls.

```
Begin
  If in a recursive call then
     N ← GN(H)
     If I(H) = 1 then Return {[H, X, Y]} End
  else
     N ← EN(H)
  End
  R ← {}
  For each side Δ of N do
     Compute q, m, l and φ_Δ
     (u, v) ← Bzout(q, m)
     For each (f, k) in Factor(L, φ_Δ) do
        ξ ← Any root of f
        H'(X, Y) ← H(ξ^v X^q, X^m(ξ^u + Y))/X^l
        For each [G, P, Q] in RNPuiseux(L(ξ), H) do
           R ← R ∪ {[G, ξ^v P^q, P^m(ξ^u + Q)]}
        End
     End
  End
```

```
    Return R
End.
```

Applied to the monic $F$, we obtain all rational Puiseux expansions above 0.

The reader may notice that rational and classical algorithms are quite similar. They differ only in the way roots of polynomials are handled and in the definition of $H'$.

In (Duval (1989)), it is suggested that the D5 system (Della Dora et al. (1985)) should be used to avoid factorization. In our case, though, since efficient algorithms are known for factoring over finite fields, factorization is not a significant issue (see Section 5).

**Example 27.** Let $F(X,Y) = (Y^2 - 2X^3)(Y^2 - 2X^2)(Y^3 - 2X) \in \mathbb{Q}[X,Y]$. Applying RNPuiseux over $\mathbb{Q}$ yields three expansions :

$$(P_1, Q_1) = (2X^2, X^0(0 + 2X^2(0 + X(2+Y)))) = (2X^2, 4X^3 + 2X^3Y)$$
$$(P_2, Q_2) = (4X^3, X^0(0 + X(2+Y))) = (4X^3, 2X + 2XY)$$
$$(P_3, Q_3) = (X, X^0(0 + X(\sqrt{2}+Y))) = (X, \sqrt{2}X + XY)$$

The first two expansions have residue field $\mathbb{Q}$ and ramification index 2 and 3. The third one corresponds to a place with residue field isomorphic to $\mathbb{Q}(\sqrt{2})$. Applying RNPuiseux over $\mathbb{Q}(\sqrt{2})$ will result in one more expansion :

$$(P_4, Q_4) = (X, X^0(0 + X(-\sqrt{2}+Y))) = (X, -\sqrt{2}X + XY).$$

**Remark 28.** It is worth noting that, unlike classical Puiseux series, rational Puiseux expansions are not canonically defined. Replacing $T$ by $\gamma T$ in $R_i(T) = (X_i(T), Y_i(T))$ with $\gamma$ chosen in the coefficient field of $R_i$ yields another rational Puiseux expansion corresponding to the same place. The choice of $\gamma_i$ can have dramatic consequences on the size of the expansion coefficients and on the performances of the algorithm. See also Section 3.6.

*3.4. Polygon trees*

To a function call RNPuiseux($L, F$) (see Section 3), we associate a labeled rooted tree. By definition, the *depth* of a vertex $v$ is the number of edges on the path from the root to $v$. In particular, the root vertex has depth 0. The tree vertices of even depth are labeled with polygons, while vertices of odd depth are labeled with integer partitions. Similarly, tree edges are labeled alternatively with edges of polygons and integer pairs $(k, f)$ where $k$ is the multiplicity of a root $\xi$ and $f = [L(\xi) : L]$. A tree edge corresponds to the choice of a polygon edge or to the choice of a characteristic polynomial root. More precisely, the tree is constructed recursively from the root vertex as follow (see Figure 3.4). Even depth vertices correspond to function calls.

- A vertex $v$ of even depth $l$ is labeled with the polygon $\mathcal{N}$, that is $\mathcal{EN}(H)$ for the root vertex ($l = 0$), and $\mathcal{GN}(H)$ for recursive calls ($l > 0$).
- To each $\Delta$ of $\mathcal{N}$ corresponds an edge from $v$ to a depth $l+1$ vertex. Label the edge with $\Delta$ (represented by its endpoint).
- A child (depth $l+1$ vertex) is labeled with the corresponding integer partition $[\phi_\Delta]$ (see the end of Section 1 for this notation).

- To each choice of root $\xi$ of $\phi_\Delta$ made by the algorithm corresponds an edge from a depth $l+1$ vertex to a depth $l+2$ vertex. The edge is labeled with the pair $(k, f)$, where $k$ is the multiplicity of $\xi$ of and $f = [L(\xi) : L]$.
- Then, we proceed recursively : A depth $l+2$ vertex is the root vertex of the tree corresponding to the function call $\texttt{RNPuiseux}(L(\xi), H')$ where $H'$ is the polynomial $H'$ obtained for a choice of edge $\Delta$ and a choice of root $\xi$.

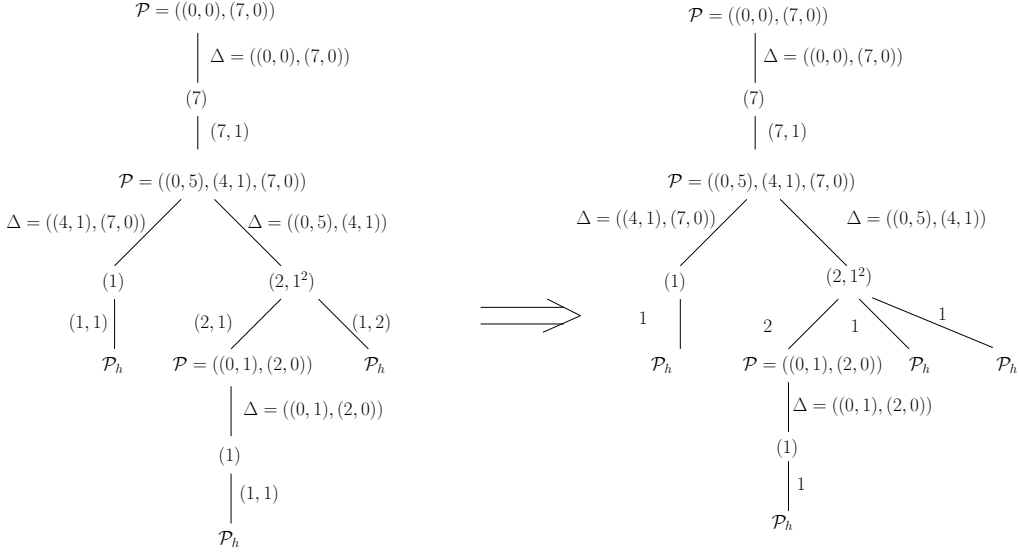

Fig. 3. Polygon trees $\mathcal{RT}(\mathbb{Q}, F)$ and $\mathcal{T}(F)$ for Example 27.

The leaves are even depth vertices labeled with polygons that have only one side $\mathcal{P}_h = [(0,1),(1,0)]$. Note that the roots $\xi$ are not part of the tree. Since the square-free factorization is a sub-product of the factorization over $L$, the labeled tree can be obtained at no significant cost. If $l$ is the depth of the function call tree generated by $\texttt{RNPuiseux}(L,F)$, then the labeled tree constructed has depth $2l$.

For a function call $\texttt{CNPuiseux}(F)$, we define a similar tree, but in this case, an edge from a partition to a polygon is only labeled with a multiplicity $k$ because the ground field is $\overline{L}$ and all field extension have degree 1.

**Definition 29.** We denote by $\mathcal{RT}(L, F)$ (resp. $\mathcal{T}(F)$) a tree associated to the function call $\texttt{RNPuiseux}(L,F)$ (resp. $\texttt{CNPuiseux}(F)$). In both cases, the tree is called *the polygon tree associated to the function call*.

It turns out that $\mathcal{T}(F)$ is precisely the symbolic information required to achieve our goal.

**Proposition 30.** *The tree $\mathcal{T}(F)$ can easily be obtained from $\mathcal{RT}(L,F)$ as follow : duplicate $f$ times each edge labeled $(k,f)$ (together with the sub-tree rooted at this edge) and replace the tag $(k,f)$ by the tag $k$.*

**Proof.** From the introduction of Section 3.5, $\mathcal{T}(F) = \mathcal{RT}(\overline{L}, F)$. Then, the construction is trivial. $\square$

This process is illustrated in Figure 3.4.

*3.5. From classical Puiseux series to rational Puiseux expansions*

Following (Duval (1989)), we remark that Newton polygons and root multiplicities obtained along the computation are the same with the two algorithms above. This remark easily extends to generic Newton polygons and associated characteristic polynomials. However, in general, the value of the nonzero roots of characteristic polynomials differ.

Studying the relations between coefficients of rational Puiseux expansions and classical coefficients has a number of benefits : it provides a better understanding of the rational algorithm, insight about the coefficient growth (see Section 3.6) and a reduction criterion for rational Puiseux expansions (see Corollary 42). Although this criterion is not strictly necessary for our algorithm, we have decided to include this material in this paper.

Let $(\alpha_1, m_1, q_1) \ldots (\alpha_h, m_h, q_h)$ be the sequence of triplets encountered in the computation of one classical Puiseux series using the `CNPuiseux` algorithm. Namely, $\alpha_i$ is a $q_i$-th root of the $i$-the characteristic polynomial and $-m_i/q_i$ is the slope of an edge of the corresponding generic Newton polygon. The output of the algorithm is :

$$P(X) = X^{q_1 q_2 \cdots q_r} = X^e$$

$$Q(X,Y) = X^{m_1 q_2 \cdots q_h}(\alpha_1 + X^{m_2 q_3 \cdots q_h}(\alpha_2 + \cdots + X^{m_h}(\alpha_h + Y)\cdots))$$

so that an element of the corresponding cycle can be written :

$$S(X) = X^{\frac{m_1}{q_1}}(\alpha_1 + X^{\frac{m_2}{q_1 q_2}}(\alpha_2 + \cdots + X^{\frac{m_h}{q_1 q_2 \cdots q_h}}(\alpha_h + \cdots)\cdots)) \tag{8}$$

On the other hand, let $(\xi_1, m_1, q_1) \ldots (\xi_r, m_h, q_h)$ be the sequence of triplets encountered in the computation of one rational Puiseux expansion using `RNPuiseux`. Now, $\xi_i$ is a root of the $i$-the characteristic polynomial and $-m_i/q_i$ is the slope of an edge of the corresponding generic Newton Polygon. We denote by $(u_i, v_i)$, $1 \leq i \leq h$ the pairs of integers returned by the `Bzout` algorithm.

Since we have used generic and exceptional Newton polygons, some of the $\xi_i$ and $\alpha_i$ may be null. If $\xi_i = \alpha_i = 0$, we have $v_i = 0$ because $\xi_i$ is associated with an edge of slope -1 or 0 and therefore, $q_i = 1$ (see procedure `Bzout`). In the sequel, we define $0^0 = 1$ so that $\xi_i^{v_i} = \alpha_i^{v_i} = 1$ and all expressions involved make sense and are correct.

**Proposition 31.** *There exists a classical Puiseux series as above and a set of integers $\{e_{ij}\}_{1 \leq j < i \leq h}$ such that :*

$$\xi_i = \alpha_i^{q_i} \prod_{j=1}^{i-1} \alpha_j^{v_j e_{ij}}.$$

**Proof.** We set $X_0 = X$ and $Y_0 = Y$, and consider transformations performed by the algorithm :

$$X_{i-1} = \xi_i^{v_i} X_i^{q_i}$$

$$Y_{i-1} = X_i^{m_i}(\xi_i^{u_i} + Y_i)$$

We define (any choice of $e$-th root is acceptable) :

$$\mu_i = \prod_{j=1}^{i} \xi_j^{-\frac{v_j}{q_j q_{j+1} \cdots q_i}} \quad 1 \leq i \leq h,$$

so that, we can write :

$$X_i = \mu_i X^{\frac{1}{q_1 q_2 \cdots q_i}}.$$

The truncated series computed by the algorithm can expressed as follow :

$$Q(X,0) = X_1^{m_1}(\xi_1^{u_1} + X_2^{m_2}(\xi_2 + X_3^{m_3}(\xi_3^{u_3} + \ldots X_{h-1}^{m_{h-1}}(\xi_{h-1}^{u_{h-1}} + X_h^{m_h}\xi_h^{u_h}) \cdots ))).$$

Using the above expression for $X_i$ and identifying coefficients with those of expression (8), there exists a classical Puiseux series so that :

$$\alpha_i \alpha_{i-1}^{-1} = \xi_{i-1}^{-u_{i-1}} \xi_i^{u_i} \mu_i^{m_i} \quad 1 \leq i \leq h$$

where we have chosen $\alpha_0 = \xi_0 = 1$. It is convenient to introduce $\theta_i = \alpha_i \alpha_{i-1}^{-1} \xi_{i-1}^{u_{i-1}}$. Hence, we have :

$$\mu_i^{q_i} = \mu_{i-1} \xi_i^{-v_i}$$
$$\theta_i = \xi_i^{u_i} \mu_i^{m_i}$$

Raising the second equality to the power $q_i$ and applying the relation $u_i q_i - m_i v_i = 1$, we obtain :

$$\xi_i = \theta_i^{q_i} \mu_{i-1}^{-m_i}.$$

Raising the first equality to the power $u_i$ and the second one to the power $v_i$, Bzout relation gives :

$$\mu_i = \theta_i^{-v_i} \mu_{i-1}^{u_i}.$$

The recurrence given by the last equality easily yields :

$$\begin{aligned}
\mu_i &= \theta_i^{-v_i} \theta_{i-1}^{-v_{i-1}u_i} \theta_{i-2}^{-v_{i-2}u_{i-1}u_i} \cdots \theta_1^{-v_1 u_2 u_3 \cdots u_i} \\
\xi_i &= \theta_i^{q_i} \left( \theta_{i-1}^{v_{i-1}} \theta_{i-2}^{v_{i-2} u_{i-1}} \cdots \theta_1^{v_1 u_2 u_3 \cdots u_{i-1}} \right)^{m_i}
\end{aligned} \quad (9)$$

Finally, the proposition is proved by induction on $i$, together with the following assertion : There exists a set of integers $\{f_{ij}\}_{1 \leq j < i \leq h}$ such that :

$$\theta_i = \alpha_i \prod_{j=1}^{i-1} \alpha_j^{v_j f_{ij}} \quad (10)$$

The case $i = 1$ is trivial. Assume that $i > 1$. The induction hypothesis about $\xi_{i-1}$ gives :

$$\theta_i = \alpha_i \alpha_{i-1}^{-1} \xi_{i-1}^{u_{i-1}} = \alpha_i \alpha_{i-1}^{-1} \alpha_{i-1}^{q_{i-1}u_{i-1}} \prod_{j=1}^{i-2} \alpha_j^{v_j e_{i-1,j} u_{i-1}}$$

Setting $f_{ij} = e_{i-1,j} u_{i-1}$ for $1 \leq j \leq i-2$ and $f_{i,j-1} = m_{i-1}$ we obtain (10). The expression for $\xi_i$ in the proposition then follows directly from the formula for $\xi_i$ in (9). □

**Remark 32.** Assuming that the $v_i$ are chosen in $\mathbb{N}$, it is easily seen that the $e_{ij}$ and $f_{ij}$ are also in $\mathbb{N}$.

**Remark 33.** Using relation (9) and the definition of $\theta_i$, it is easy to express recursively the $\xi_i$ in terms of the $\alpha_i$, but there is no simple formula. On the other hand, the $\alpha_i$ can be easily expressed as follow : For $1 \leq i \leq j \leq h$, define $s_{ji} = \sum_{k=j}^{i} \frac{m_k}{q_j \cdots q_k}$. Then :

$$\alpha_i = \xi_i^{u_i} \prod_{j=1}^{i} \xi_j^{-v_j s_{ji}}.$$

To conclude this part, we rewrite the coefficients of parametrizations returned by `RNPuiseux` in terms of the $\xi_i$. In order to simplify expressions, we introduce the following notation for $0 \leq i \leq h-1$ :

$$\xi_{(i)} = \xi_{i+1}^{v_{i+1}} \xi_{i+2}^{v_{i+2} q_{i+1}} \xi_{i+3}^{v_{i+3} q_{i+1} q_{i+2}} \cdots \xi_h^{v_h q_{i+1} \cdots q_{h-1}}.$$

We also define $\xi_{(h)} = 1$. The output is now given by :

$$P(X) = \xi_{(0)} X^{q_1 q_2 \cdots q_h} = \xi_{(0)} X^e$$
$$Q(X,Y) = \xi_{(1)}^{m_1} X^{\frac{m_1}{q_1}} (\xi_1^{u_1} + \xi_{(2)}^{m_2} X^{\frac{m_2}{q_1 q_2}} (\xi_2^{u_2} + \cdots + \xi_{(h)}^{m_h} X^{\frac{m_h}{q_1 q_2 \cdots q_h}} (\xi_h^{u_h} + Y) \cdots)).$$

We deduce the parametrization :

$$\begin{aligned}
X(T) &= \xi_{(0)} T^e \\
Y(T) &= \xi_1^{u_1} \xi_{(1)}^{m_1} T^{m_1 q_2 \cdots q_h} + \\
&\quad \xi_2^{u_2} \xi_{(1)}^{m_1} \xi_{(2)}^{m_2} T^{m_1 q_2 \cdots q_h + m_2 q_3 \cdots q_h} + \\
&\quad \cdots \\
&\quad \xi_h^{u_h} \xi_{(1)}^{m_1} \xi_{(2)}^{m_2} \cdots \xi_{(h)}^{m_h} T^{m_1 q_2 \cdots q_h + m_2 q_3 \cdots q_h + \cdots + m_h} + \cdots
\end{aligned} \quad (11)$$

*3.6. Rational Puiseux expansion coefficient bit-size*

In (Walsh (1999)), it is mentioned without evidence that, unlike classical Puiseux series, rational Puiseux expansions can have coefficients with exponential bit-size. Since we shall compute rational Puiseux expansions only over finite fields, the coefficient bit-size is essentially bounded by the logarithm of the residue field cardinal. Therefore, this bad behavior is not a significant drawback for us.

**Example 34.** We illustrate the coefficient growth of rational Puiseux expansions and the impact of the choice of $u$ and $v$ in the procedure `Bzout` with the following example. The growth is not exponential, but still, leads to dramatically large coefficients. Let $a$ and $h$ be positive integers and consider the following parametrization, introduced in a different context (Henry and Merle (1987)) :

$$X(T) = T^{2^h}, Y(T) = \sum_{k=1}^{h} a\, T^{3\, 2^h (1-1/2^k)} \quad (12)$$

We define $d = 2^h$ and $F_d(X,Y) = \text{Resultant}_T(X - X(T), Y - Y(T))$, so that $\deg_Y(F_d) = d$ and $\deg_X(F_d) = 3(d-1)$. There is a unique place above 0 for $F_d$ and therefore, a system of rational Puiseux expansion contains a unique parametrization. We use the notations of the previous subsection. Applying `RNPuiseux` to $F_d$, we see that $h$ is the number of recursive calls and $m_i = 3$, $q_i = 2$ for $1 \leq i \leq h$.

- **First strategy.** We choose $v_i$ nonnegative and minimal, that is $u_i = 2$ and $v_i = 1$. We have $\theta_1 = a$ and, for $i > 1$, $\theta_i = \alpha_i \alpha_{i-1}^{-1} \xi_{i-1}^2 = \xi_{i-1}^2$ so that relation (9) yields:

$$\xi_i = \xi_{i-1}^4 (\xi_{i-1}^2 \xi_{i-3}^4 \ldots \xi_1^{2^{i-2}} a^{2^{i-2}})^3.$$

Therefore, $\xi_i$ is an integer power of $a$ and crude estimates show that the exponent is greater than $4^i$ for $i > 3$. Substituting in $\xi_{(0)}$, we deduce that $\xi_{(0)}$ is a power of $a$ with exponent larger than $4^h 2^{h-1} > d^3/2$ for $h > 3$. Hence, the bit-size of $\xi_{(0)}$ is (much) larger than $d^3/2 \log(a)$ for $h > 3$. The behavior is clearly worse for the other coefficients of (11). For $F_{16}$, RNPuiseux returns :

$$X(T) = a^{3072} T^{16}$$
$$Y(T) = a^{4609} T^{24} + a^{6913} T^{36} + a^{8065} T^{42} + a^{8641} T^{45}$$

Considering (12), this result is far from optimal !

- **Second strategy.** Another reasonable choice is $u_i = v_i = -1$. Hopefully, this will result in smaller coefficients since the exponents in (9) will be considerably smaller and have alternate signs. For $F_{16}$, we obtain :

$$X(T) = a^{528} T^{16}$$
$$Y(T) = a^{793} T^{24} + a^{1189} T^{36} + a^{1387} T^{42} + a^{1486} T^{45}$$

Again, we fall short from a satisfactory result.

- **Maple 10 strategy.** The algcurves[puiseux] command returns an even worse result :

$$X(T) = a^{24672} T^{16}$$
$$Y(T) = a^{37009} T^{24} + a^{55513} T^{36} + a^{64765} T^{42} + a^{69391} T^{45}$$

As remarked in Duval (1989), optimal output is not always reachable by the current algorithm, no matter how $u$ and $v$ are chosen, even for simple cases. Transformations different than that of RNPuiseux may be necessary. In this example, reduction of powers of $a$ in the course of the computation by substitutions of the form $T = U/a^s$ may result in smaller coefficients, but it is not clear how efficient this workaround may be in general.

If $L$ is a number field, Walsh has shown that a coefficient $\gamma$ with bit-size in $O(d^{11} n^3)$ can be constructed so that $X(T) = \gamma T^e$ and all other expansion coefficients have polynomial bit-size (Walsh (1999)). Walsh's construction requires that a Puiseux series is already known and does not allow to optimize the size in the course of the algorithm. It is not clear how far this construction is from an optimal size.

In summary, an algorithm that computes rational Puiseux expansions with provably small coefficient size is still unknown.

## 4. Good reduction

We consider a polynomial $F$ with coefficients in an algebraic number field $K$ and discuss how to choose a prime number $p$ so that the computation of rational Puiseux expansions modulo $p$ provides enough information to guide floating point computations of Puiseux series, namely $\mathcal{T}(F)$.

We denote by $\mathfrak{o}$ the ring of algebraic integers of $K$. If $\mathfrak{p}$ is a prime ideal of $\mathfrak{o}$, $v_\mathfrak{p}$ the corresponding valuation of $K$. Finally, we define :

$$\mathfrak{o}_\mathfrak{p} = \{\alpha \in K \,|\, v_\mathfrak{p}(\alpha) \geq 0\}.$$

Let $L$ be the finite extension generated over $K$ by the Puiseux series coefficients of $F$. Note that by Proposition 31, $L$ also contains the coefficients of rational Puiseux expansions computed by RNPuiseux. If $\mathfrak{O}$ stands for the ring of algebraic integers of $L$ and $\mathfrak{P}$ a prime ideal of $\mathfrak{O}$, we introduce :

$$\mathfrak{O}_\mathfrak{P} = \{\alpha \in L \,|\, v_\mathfrak{P}(\alpha) \geq 0\}.$$

In the sequel, $\mathfrak{P}$ will always denote a prime ideal of $\mathfrak{O}$ dividing $\mathfrak{p}$.

The reduction modulo $\mathfrak{P}$ of $\alpha \in \mathfrak{O}_\mathfrak{P}$ is represented by $\overline{\alpha}$. We extend this notation to polynomials and fractional power series with coefficients in $\mathfrak{O}_\mathfrak{P}$. If $\alpha \in \mathfrak{o}_\mathfrak{p}$, since $\mathfrak{P}$ divides $\mathfrak{p}$, reduction modulo $\mathfrak{P}$ and $\mathfrak{p}$ coincide and we shall use the same notation $\overline{\alpha}$.

*4.1. Modular reduction of Puiseux series*

Our reduction strategy is based on the following definition :

**Definition 35.** Let $p$ be a prime number and $\mathfrak{p}$ a prime ideal of $\mathfrak{o}$ dividing $p$. If the conditions below are verified :
- $F \in \mathfrak{o}_\mathfrak{p}[X, Y]$,
- $p > d$,
- $v_\mathfrak{p}(\text{tc}(\Delta_F)) = 0$.

we say that $F$ has *good reduction at* $\mathfrak{p}$.

Note that if $F$ has good reduction at $\mathfrak{p}$, since $\mathfrak{P}$ divides $\mathfrak{p}$, $v_\mathfrak{P}(\text{tc}(\Delta_F)) = 0$ and $F$ has also good reduction at $\mathfrak{P}$. We shall use this fact freely in the sequel.

A fundamental result for the reduction strategy is the following consequence of a theorem by Dwork and Robba :

**Theorem 36.** *If $F$ has good reduction at $\mathfrak{p}$, then the Puiseux series coefficients of $F$ above 0 are in $\mathfrak{O}_\mathfrak{P}$.*

**Proof.** Let $S(X) = \sum_{i=0}^\infty \alpha_i X^{i/e}$ be any of the $S_{ij}(X)$. We write $\mathbb{C}_p$ for the field of $p$-adic numbers, that is the completion of the algebraic closure of $\mathbb{Q}_p$ (completion of $\mathbb{Q}$ for $p$-adic absolute value). Consider the field $L$ as a subfield of $\mathbb{C}_p$ by means of its $\mathfrak{P}$-adic completion. We denote by $|.|_p$ the absolute value of $\mathbb{C}_p$, so that :

$$\mathfrak{O}_\mathfrak{P} = \{\alpha \in L \,|\, |\alpha|_p \leq 1\}.$$

Since the coefficients of $\Delta_F$ are in $\mathfrak{O}_\mathfrak{P}$ and $|\text{tc}(\Delta_F)|_p = 1$, $\Delta_F$ has no nonzero root in $D(0, 1^-) = \{\alpha \in \mathbb{C}_p \,|\, |\alpha|_p < 1\}$. Therefore, Theorem 2.1 of (Dwork and Robba (1979)) asserts that the $p$-adic convergence radius of $S(X)$ is at least 1. Moreover, $S(X)$ is bounded by 1 on $D(0, 1^-)$, otherwise, $S(x)$ cannot satisfy a monic equation $S(x)^d + A_{d-1}(x)S(x)^{d-1} + \cdots + A_0(x) = 0$ for $x \in D(0, 1^-)$.

A classical isometry of $p$-adic analysis (see Robert (2000), Section 4.6) :
$$\sup_{x \in D(0,1^-)} |S(x)|_p = \sup_{i \geq 0} |\alpha_i|_p$$
yields $|\alpha_i|_p \leq 1$ so that $\alpha_i \in \mathfrak{O}_\mathfrak{P}$. □

It is worth insisting on the fact that this result holds for *any* $\mathfrak{P}$ dividing $\mathfrak{p}$.

**Example 37.** Consider the case $F(X,Y) = Y^2 - X^3(p+X)$ with $p > 2$. Puiseux series above 0 are :
$$S_{1j}(X) = (-1)^j \sqrt{p} X^{3/2}(1 + \frac{X}{p})^{1/2} = (-1)^j \sqrt{p} X^{3/2}(1 + \frac{X}{2p} - \frac{X^2}{8p^2} + \cdots).$$
They are obviously not reducible modulo $p$, but the discriminant criteria of the theorem detects this deficiency. It is interesting to note, however, that a system of rational Puiseux expansions is given by $\{X = pT^2, Y = p^2 T^3 + \frac{1}{2} p^2 T^5 + \dots\}$. This parametrization is reducible modulo $p$, but the reduction $\{X = 0, Y = 0\}$ is trivial and hardly useful. On the other hand, $\{X = T^2/p, Y = T^3/p + \frac{1}{2} T^5/p^3 + \dots\}$ is also a (non reducible) system of rational Puiseux expansions.

**Remark 38.** This reduction criterion could also be deduced from (Dwork (1984)). Under the hypotheses of Theorem 36, it is shown therein that Puiseux series modulo $p$ can be lifted to Puiseux series with coefficients in $\mathbb{C}_p$. The latter must coincide with the embedding in $\mathbb{C}_p$ of Puiseux series over $L$. Therefore, Puiseux series over $L$ can be reduced modulo a prime ideal dividing $p$.

**Corollary 39.** *If $F$ has good reduction at $\mathfrak{p}$, then any monic factor $G$ of $F$ in $\overline{K}[[X]][Y]$ satisfies $v_\mathfrak{P}(\text{tc}(\Delta_G)) = 0$.*

**Proof.** Let $F = GH$. Both $G$ and $H$ are monic, so relation (2) shows that $\text{tc}(\Delta_F) = \text{tc}(\Delta_G)\text{tc}(\Delta_H)\text{tc}(\text{Resultant}(G,H))$. From Theorem 36, the coefficients of $G$ and $H$ are in $\mathfrak{O}_\mathfrak{P}$, and so are these numbers. The result is then a trivial consequence of $v_\mathfrak{p}(\text{tc}(\Delta_F)) = v_\mathfrak{P}(\text{tc}(\Delta_F)) = 0$. □

**Corollary 40.** *If $F$ has good reduction at $\mathfrak{p}$, then characteristic coefficients of all ramified Puiseux series of $F$ above 0 have $\mathfrak{P}$-adic valuation zero. In other words, reduction modulo $\mathfrak{P}$ preserves the characteristic of ramified cycles of $F$ above 0.*

**Proof.** Follows easily from Corollary 39 applied to the irreducible factors of $F$ in $\overline{K}[[X]][Y]$ and from Proposition 6. □

It is important to note, however, that annihilation modulo $\mathfrak{P}$ of Puiseux series coefficients is not totally controlled by our good reduction criterion. If $F$ is irreducible in $\overline{K}[[X]][Y]$, all non-characteristic coefficients may vanish modulo $\mathfrak{P}$, as shown by Proposition 6 (consider for instance the minimal polynomial over $\mathbb{Q}(X)$ of $S(X) = pX + X^{3/2}$). If $F$ is not irreducible, our criterion will also detect cancellation of coefficients that "separate" cycles. This property is contained in Theorem 43.

**Theorem 41.** *Let $\{S_i\}_{1\leq i\leq s}$ be a set of representatives for the cycles of $F$ above $0$. Assume that $F$ has good reduction at $\mathfrak{p}$. Then, $\{\overline{S_i}\}_{1\leq i\leq s}$ form a set of representatives for the cycles of $\overline{F}$ above $0$.*

**Proof.** The $\overline{S_i}$ satisfy $\overline{F}(X, \overline{S_i}) = 0$ and $\Delta_{\overline{F}} = \overline{\Delta_F} \neq 0$. Therefore, the $\overline{S_i}$ are distinct roots of $\overline{F}$. By Corollary 40, the ramification index of $\overline{S_i}$ is equal to the ramification index of $S_i$, namely $e_i$. Since $\sum_{i=1}^{s} e_i = d$, we have obtained a complete set of representatives for the cycles of $\overline{F}$. □

We now show that parametrizations computed by `RNPuiseux` yield meaningful parametrizations when reduced modulo $\mathfrak{P}$ (see Example 37).

**Corollary 42.** *Denote by $R(T) = (\gamma T^e, \sum_{i=0}^{r} \beta_i T^{a_i})$ (with $\beta_i \neq 0$) a parametrization given by `RNPuiseux`. If $F$ has good reduction at $\mathfrak{p}$, then the $\beta_i$ belong to $\mathfrak{O}_{\mathfrak{P}}$ and $v_{\mathfrak{P}}(\gamma) = 0$.*

**Proof.** We use the notations of Section 3.5. If $\alpha_j$ is a characteristic coefficient, then Corollary 40 shows that $v_{\mathfrak{P}}(\alpha_j) = 0$. If $\alpha_j$ is not a characteristic coefficient, it is the root of a characteristic polynomial of a Newton polygon edge with integer slope. Hence, $q_j = 1$, $v_j = 0$ and $u_j = 1$ (see procedure `Bzout`). From Proposition 31 we deduce that $v_{\mathfrak{P}}(\xi_i) = q_i v_{\mathfrak{P}}(\alpha_i)$ for all $1 \leq i \leq h$. The same argument proves that $v_{\mathfrak{P}}(\xi_{(i)}) = 0$ for $0 \leq i \leq h$. In particular, $\gamma = \xi_{(0)}$ and $v_{\mathfrak{P}}(\gamma) = 0$. Finally, (11) shows that $v_{\mathfrak{P}}(\beta_i) = u_i q_i v_{\mathfrak{P}}(\alpha_i)$. If $\alpha_i$ is a characteristic coefficient, the latter valuation is zero, otherwise it is equal to $q_i v_{\mathfrak{P}}(\alpha_i) \geq 0$ since $u_i = 1$ in this case. □

*4.2. Modular reduction of polygon trees*

If $F \in \mathfrak{o}_{\mathfrak{p}}[X, Y]$ and $p > d$, algorithms of Section 3 can be applied to the reduction $\overline{F}$ of $F$ modulo $\mathfrak{p}$, so that the notations $\mathcal{T}(\overline{F})$ and $\mathcal{RT}(\mathbb{F}_{p^t}, \overline{F})$ make sense. The computed expansions have coefficients in a finite extension of $\mathbb{F}_p$.

The following result is crucial. It allows to compute by means of modular computations the symbolic information required :

**Theorem 43.** *If $F$ has good reduction at $\mathfrak{p}$, then $\mathcal{T}(F) = \mathcal{T}(\overline{F})$.*

Note that the correspondence between $\mathcal{T}(F)$ and $\mathcal{T}(\overline{F})$ cannot be stated so simply if classical Newton polygons are used instead of generic ones : Non-characteristic coefficients of Puiseux series may vanish upon modular reduction, yielding polygon modifications.

We begin with a number of lemmas :

**Lemma 44.** *Assume that $H$ satisfies :*
  (i) $H \in \mathfrak{O}_{\mathfrak{P}}[X, Y]$,
 (ii) *$H$ has no multiple roots, $\deg_Y H = d > 0$, $H(0, 0) = 0$, $H(0, Y) \neq 0$,*
(iii) *the leading coefficient of $H$ as a polynomial in $Y$ is a power of $X$,*
(iv) *the roots of $H$ are in $\cup_{e>0} \mathfrak{O}_{\mathfrak{P}}((X^{1/e}))$,*
 (v) *$\overline{H}$ has no multiple roots.*
(vi) $v_{\mathfrak{P}}(\text{tc}(\Delta_H)) = 0$.

Let $(m, q, l)$ be integers associated to an edge $\Delta$ of $\mathcal{GN}(H)$ and let $\xi$ be a root of $\phi_\Delta$. Then, $H'(X, Y) = H(X^q, X^m(\xi + Y))/X^l$ also satisfies the conditions (i) to (vi).

**Proof.** Conditions $H'(0,0) = 0$ and $H'(0,Y) \neq 0$ follow from the properties of `CNPuiseux`. The leading coefficient property is obviously preserved by the transformation. If $\{Y_i(X)\}_{1 \leq i \leq d}$ denotes the roots of $H$, the roots of $H'$ are $\{Y_i(X)/X^m - \xi\}_{1 \leq i \leq d}$. It is clear that they are distinct. Since $\xi$ is in $\mathfrak{O}_\mathfrak{P}$, so are the coefficients of $H'$ and of its roots. Finally, since the discriminant is, up to a power of $X$, a product of root differences, its trailing coefficient does not change under the transformation. $\square$

**Lemma 45.** *Assume that $H$ satisfies the condition of Lemma 44. Then:*
  *(i) $\mathcal{GN}(H) = \mathcal{GN}(\overline{H})$.*
  *(ii) Let $\Delta$ be a edge of $\mathcal{GN}(H)$. The characteristic polynomials $\phi_\Delta$ (resp. $\overline{\phi_\Delta}$) of $\Delta$ in $H$ (resp. $\overline{H}$) satisfy $[\phi_\Delta] = [\overline{\phi_\Delta}]$ (equality of root multiplicities).*

**Proof.** Denote $\{S_i\}_{1 \leq i \leq w}$ the cycles of $H$ that vanish at $X = 0$ and $\{H_i\}_{1 \leq i \leq w}$ their minimal polynomials over $\overline{K}((X))$. The assumption about the roots of $H$ induces that $H_i$ belongs to $\mathfrak{O}_\mathfrak{P}[[X]][Y]$. We define $V = \prod_{i=1}^{w} H_i$, so that $V$ is a monic polynomial with coefficient in $\mathfrak{O}_\mathfrak{P}$. Define $U$ so that $H = UV$.

By Proposition 25, $\mathcal{I}(H) = \mathcal{I}(V)$, thus $U(0,0) \neq 0$. Necessarily, $\mathcal{GN}(H) = \mathcal{GN}(V)$. We first show that $\mathcal{GN}(\overline{H}) = \mathcal{GN}(\overline{V})$, which is equivalent to $v_\mathfrak{P}(U(0,0)) = 0$. Since the leading coefficient of $H$ is a power of $X$, relation (2) shows that:

$$\operatorname{tc}(\Delta_H) = \operatorname{tc}(\Delta_V)\operatorname{tc}(\Delta_U)\operatorname{tc}(\operatorname{Resultant}(U,V))^2.$$

But $V$ is monic, so the latter resultant is $\pm \prod_i U(X, v_i)$, where $v_i$ runs through the roots of $V$. Since $v_i(0) = 0$, it is obvious that $\operatorname{tc}(\Delta_H)$ is the product of a power of $U(0,0)$ and of an element of $\mathfrak{O}_\mathfrak{P}$. The hypothesis $v_\mathfrak{P}(\operatorname{tc}(\Delta_H)) = 0$ yields $v_\mathfrak{P}(U(0,0)) = 0$. Thus, $\mathcal{GN}(\overline{H}) = \mathcal{GN}(\overline{V})$.

To prove *(i)*, by Lemma 21, it remains to show that $\mathcal{GN}(H_i) = \mathcal{GN}(\overline{H_i})$. If $\mathcal{I}(H_i) = 1$, then $\mathcal{I}(\overline{H_i}) = 1$ because $H_i$ is monic. Therefore, both $\mathcal{GN}(H_i)$ and $\mathcal{GN}(\overline{H_i})$ are reduced to the unique edge $[(0,1), (1,0)]$. Assume that $\mathcal{I}(H_i) > 1$. It is easily seen that the hypotheses imply $v_\mathfrak{P}(\operatorname{tc}(\Delta_{H_i})) = 0$. Proposition 6 shows that $S_i$ and $\overline{S_i}$ have the same characteristic. In particular, $\overline{H_i}$ is irreducible in $\overline{\mathbb{F}_p}[[X]][Y]$ and $\mathcal{GN}(\overline{H_i})$ has a single edge, whose characteristic polynomial has a unique root. If the unique edge of $\mathcal{GN}(H_i)$ has slope $-1$, so does the unique edge of $\mathcal{GN}(\overline{H_i})$ since the vanishing modulo $\mathfrak{P}$ of $\operatorname{tc}(H_i(X, 0))$ leads to the same (fictitious) edge. If the unique edge has a slope greater than $-1$, $\operatorname{tc}(H_i(X, 0))$ is a nonnegative power of a characteristic coefficient and does not vanish modulo $\mathfrak{P}$. In both cases, $\mathcal{GN}(H_i) = \mathcal{GN}(\overline{H_i})$.

To address *(ii)*, let $\Delta$ be a common edge of $\mathcal{GN}(H)$ and $\mathcal{GN}(\overline{H})$. If $\Delta$ corresponds to unique irreducible $H_i$ and $\overline{H_i}$, $\phi_\Delta$ and $\overline{\phi_\Delta}$ have a unique root with the same multiplicity, since they have the same degree, and we are done. Assume that $\Delta$ corresponds to a least two irreducible polynomials $H_1$ and $H_2$ associated to the roots $\xi_1$ and $\xi_2$ of $\phi_\Delta$. In order to demonstrate *(ii)*, we just need to show that if $\xi_1 \neq \xi_2$, then $\overline{\xi_1} \neq \overline{\xi_2}$. If $m$ and $q$ are relatively prime integers such that $-m/q$ is the slope of $\Delta$, we set $\alpha_i = \xi_i^{1/q}$ (any choice of $q$-root suits). The cycle associated to $H_i$ can be represented by the series $\alpha_i X^{m/q} + \cdots$. By (1), there exists $\delta \in \mathfrak{O}_\mathfrak{P}$ such that $\operatorname{tc}(\Delta_H) = (\alpha_1 - \alpha_2)\delta$. Hence, $v_\mathfrak{P}(\alpha_1 - \alpha_2) = 0$, so that $\overline{\alpha_1} \neq \overline{\alpha_2}$, which in turn gives $\overline{\xi_1} \neq \overline{\xi_2}$. $\square$

**Lemma 46.** *Assume that $F$ has good reduction at $\mathfrak{p}$. Then :*
  (i) $\mathcal{EN}(F) = \mathcal{EN}(\overline{F})$.
  (ii) *Let $\Delta$ be the unique edge of $\mathcal{EN}(F)$. The characteristic polynomial $\phi_\Delta$ (resp. $\overline{\phi_\Delta}$) of $\Delta$ in $F$ (resp. $\overline{F}$) satisfy $[\phi_\Delta] = [\overline{\phi_\Delta}]$.*

**Proof.** Since $F$ is monic, $(i)$ is trivially correct by definition of $\mathcal{EN}(F)$. As for $(ii)$, remark that $\phi_\Delta = F(0,T)$ (resp. $\overline{\phi_\Delta} = \overline{F(0,T)}$) and that the $d$ roots of $r(0,T)$ (resp. $\overline{F(0,T)}$) are precisely the $d$ values of $S_{ij}(0)$ (resp. $\overline{S_{ij}(0)}$). The good reduction hypothesis implies by (1) that if $S_{ij}(0) \neq S_{uv}(0)$, then $\overline{S_{ij}}(0) \neq \overline{S_{uv}}(0)$. Therefore, the multiplicities of the roots of $\phi_\Delta$ are preserved by reduction modulo $\mathfrak{P}$. A fortiori, the squarefree factorization of $\phi_\Delta$ is preserved modulo $\mathfrak{p}$. □

**Remark 47.** The assertion $(i)$ of Lemma 46 does not hold if the exceptional polygon is replaced by the generic polygon, as shown by the example $F(X,Y) = (Y + p + X)(Y + 1 + X)$. The good reduction criterion does not detect the cancellation of $F(0,0)$, but does detect a change of root multiplicities. This remark justifies the introduction of $\mathcal{EN}(F)$.

**Proof.** (of Theorem 43) By Lemma 46, the root vertex, depth one vertices and edges down to depth 2 vertices of $\mathcal{T}(\overline{F})$ are correctly labelled.

Let $\xi$ be a root of $F(0,Y)$ and $H(X,Y) = F(X, Y+\xi)$. Assumptions of Lemma 44 are obviously satisfied for $H$ because $\xi \in \mathfrak{O}_\mathfrak{P}$ and $\Delta_H = \Delta_F$. Denote by $\mathcal{T}'(H)$ the sub-tree of $\mathcal{T}(F)$ corresponding to the recursive function call CNPuiseux($H$).

We show that, for all $H$ satisfying hypotheses of 44, $\mathcal{T}'(H) = \mathcal{T}'(\overline{H})$. We proceed by induction on the number $c$ of function calls to CNPuiseux necessary to compute $\mathcal{T}'(H)$.

If $c = 1$, then $\mathcal{I}(H) = 1$ and $\mathcal{T}'(H)$ is reduced to a single vertex labelled with $\mathcal{GN}(H)$, which consists of the unique edge $[(0,1),(1,0)]$. Lemma 45 gives $\mathcal{T}(H) = \mathcal{T}(\overline{H})$.

Suppose now that $c > 1$. Lemma 45 shows that the root vertex, the depth 1 vertices and all edges from the root to depth 2 vertices of $\mathcal{T}'(H)$ and $\mathcal{T}'(\overline{H})$ coincide and are labelled identically. Let $H'$ denote a polynomial obtained from $H$ in CNPuiseux. The number of function calls necessary to compute $\mathcal{T}'(H')$ is less than $c$. Lemma 44 ensures that the induction hypotheses can be applied to $H'$. Hence, $\mathcal{T}'(H') = \mathcal{T}'(\overline{H'})$. By construction of polygon trees, $\mathcal{T}'(H) = \mathcal{T}'(\overline{H})$  □

*4.3. Choosing a good prime*

This part is devoted to the choice of a prime ideal $\mathfrak{p}$ such that $F$ has good reduction at $\mathfrak{p}$.

Assume that $K = \mathbb{Q}(\gamma)$ and let $M_\gamma$ be the minimal polynomial of $\gamma$ over $\mathbb{Q}$. The elements of $K$ are represented as polynomials in $\gamma$ of degree less than $w = [K : \mathbb{Q}]$ with coefficients in $\mathbb{Q}$. Up to a change of variable in $M_\gamma$ and the coefficients of $F$, we suppose that $\gamma$ belongs to $\mathfrak{o}$, namely $M_\gamma \in \mathbb{Z}[T]$.

**Definition 48.** Let $P$ be a multivariate polynomial of $K[\underline{X}]$. There exist a unique pair $(H, c)$ with $H \in \mathbb{Z}[T, \underline{X}]$, $c \in \mathbb{N}$, $\deg_T < w$ and $P(\underline{X}) = H(\gamma, \underline{X})/c$, where $c$ is minimal. The polynomial $H$ is called the *numerator of $P$* and is denoted num$(P)$. The integer $c$ is called the *denominator* of $H$ and is written denom$(P)$. We define the *size of $P$* as follow : ht$(P) = \max\{\log c, \log \|R\|_\infty\}$.

Defining $F_n = \mathrm{num}(F)$ and $b = \mathrm{denom}(F)$, we have :

$$F(X,Y) = \frac{F_n(\gamma, X, Y)}{b}.$$

We are left with the problem of finding a prime number $p$ and a prime ideal $\mathfrak{p}$ of $\mathfrak{o}$ dividing $p$ such that :

$(C_1)$ $p > d$.

$(C_2)$ $p$ does not divide $b$.

$(C_3)$ We can determine an explicit representation of a prime ideal $\mathfrak{p}$ of $\mathfrak{o}$ dividing $p$, so that a morphism $\mathfrak{o} \to \mathfrak{o}/\mathfrak{p} \cong \mathbb{F}_{p^t}$ can be effectively computed.

$(C_4)$ $\mathrm{tc}(\Delta_F) \not\equiv 0$ modulo $\mathfrak{p}$.

Condition $(C_1)$ and $(C_2)$ are easily verified. We deal with condition $(C_3)$ in a standard fashion. Let $\overline{M}$ be any irreducible factor of $M_\gamma$ in $\mathbb{F}_p[T]$ and $M$ be a lifting of $\overline{M}$ in $\mathbb{Z}[T]$. It is well-known that if $p$ is a prime number not dividing the index $e_\gamma = [\mathfrak{o} : \mathbb{Z}(\gamma)]$ and if $\overline{M}$ is any irreducible factor of $M_\gamma$ in $\mathbb{F}_p[T]$, then the ideal $\mathfrak{p} = (p, M(\gamma))$ of $\mathfrak{o}$ is prime (Cohen (1993)). Hence, elements of $\mathfrak{o}$ can be reduced by means of the morphism $\mathfrak{o} \to \mathfrak{o}/\mathfrak{p} \cong \mathbb{F}_p[T]/(\overline{M}) \cong F_{p^t}$ where $t = \deg \overline{M}$. Computing $e_\gamma$ is a non-trivial task, and so is the computation of generators of prime ideals dividing $p$ when $p$ divides $e_\gamma$. If $e_\gamma$ is unknown, it is sufficient to choose $p$ so that it does not divide $\Delta_{M_\gamma}$, since $e_\gamma$ divides $\Delta_{M_\gamma}$.

In practice, $\overline{M}$ is chosen amongst the factors of $\overline{M_\gamma}$ of smallest degree. Moreover, it is worth trying a few primes $p$ in order to reduce $t$, the case $t = 1$ being the most favorable.

As for $(C_4)$, deterministic and randomized strategies are studied in the next subsections. In order simplify the analysis, we replace condition $(C_4)$ by the following stronger condition :

$(C'_4)$ $\mathrm{Norm}_{K/\mathbb{Q}}(\mathrm{tc}(\Delta_F)) \not\equiv 0$ modulo $p$.

If $(C_1)$ to $(C'_4)$ are verified, then for all prime ideals $\mathfrak{p}$ dividing $p$, $F$ has good reduction at $\mathfrak{p}$. In practice, though, we do not recommend to use $(C'_4)$. Finally, we introduce the notation :

$$N_F = b|\mathrm{Norm}_{K/\mathbb{Q}}(\mathrm{tc}(\Delta_F))\Delta_{M_\gamma}|.$$

Conditions $(C_1)$ to $(C'_4)$ are equivalent to :

$(C_5)$ $p > d$ and $N_F \not\equiv 0$ modulo $p$.

*4.3.1. Deterministic strategy*

We determine a bound $B$ such that, for all prime numbers $p > B$, condition $(C_5)$ is satisfied. We prove first two lemmas that will also be useful in the next section.

**Lemma 49.** *The discriminant $\Delta_{F_n} \in \mathbb{Z}[X,T]$ of $F_n$ with respect to $Y$ satisfies :*

$$||\Delta_{F_n}||_\infty \leq (2d-1)!\, d^d\,[(w+1)(n+1)]^{2d-2}\,||F_n||_\infty^{2d-1}.$$

**Proof.** Expanding the discriminant of the Sylvester matrix of $F_n$ and $F_{nY}$, we see that there exist $d-1$ coefficients $\{a_i\}_{1 \leq i \leq d-1}$ of $F_n$ and $d$ coefficients $\{a_{i+d-1}\}_{1 \leq i \leq d}$ of $F_{nY}/d$ such that :

$$||\Delta_{F_n}||_\infty \leq (2d-1)!\, d^d\, ||\prod_{k=1}^{2d-1} a_i(X,T)||_\infty,$$

The bound follows recursively from :

$$||a_i c||_\infty \leq (w+1)(n+1)||a_i||_\infty ||c||_\infty$$

for all $c(X,T)$ and the inequality $||a_i||_\infty \leq ||F_n||_\infty$. □

We denote by $R_F(T)$ the numerator of $\mathrm{tc}(\Delta_F)$. Note that $\mathrm{denom}(\mathrm{tc}(\Delta_F))$ is a power of $b$ dividing $b^{2d-1}$.

**Lemma 50.**

$$||R_F||_\infty \leq ||\Delta_{F_n}||_\infty (||M_\gamma||_\infty + 1)^{(w-1)(2d-1)-w+1} = B_0 \qquad (13)$$

$$|\mathrm{Norm}_{K/\mathbb{Q}}(R_F(\gamma))| \leq (w+1)^{(2w-1)/2} ||M_\gamma||_\infty^{w-1} B_0^w = B_1 \qquad (14)$$

$$e_\gamma \leq |\Delta_{M_\gamma}| \leq w^w (w+1)^{(2w-1)/2} ||M_\gamma||_\infty^{2w-1} = B_2. \qquad (15)$$

**Proof.** Since the leading coefficient of $F_n$ in $Y$ is a constant, evaluation and resultant commute and we have :

$$b^{2d-1}\Delta_F(X) = \Delta_{F_n(\gamma,X,Y)} = \Delta_{F_n}(\gamma, X).$$

Let $C_i(T) \in \mathbb{Z}[T]$ be the coefficient of $X^i$ in $\Delta_{F_n}(T,X)$ and $c_i(T) \in \mathbb{Z}[T]$ be the numerator of the coefficient of $X^i$ in $\Delta_{F_n(\gamma,X)}$. It is clear that $c_i(\gamma) = C_i(\gamma)$. Since $M_\gamma$ is monic, Euclidean division yields $Q_i \in \mathbb{Z}[T]$ such that $C_i = Q_i M_\gamma + c_i$. Since $\deg_T C_i \leq (2d-1)(w-1)$, one can show that :

$$||c_i||_\infty \leq ||C_i||_\infty (||M_\gamma||_\infty + 1)^{(w-1)(2d-1)-w+1}.$$

The latter inequality gives (13). Now, from $\mathrm{Norm}_{K/\mathbb{Q}}(R_F(\gamma)) = \mathrm{Resultant}(M, R_F)$, Hadamard's inequality and trivial comparison of norms yield (14). The same arguments show that (15) holds. □

Finally, we have the following result, for which we do not claim optimality :

**Proposition 51.** *Define $B = \max\{b, d+1, B_1, B_2\}$ (see (14) and (15)). Then, for all $p > B$ condition $(C_5)$ is verified. Moreover, $B$ is effectively computable and there exists a prime $p > B$ with size $\mathrm{ht}(p) \in O(wd(w\,\mathrm{ht}(M_\gamma) + \mathrm{ht}(F) + \log(wnd)))$.*

**Proof.** If $p$ is a prime greater than $B$, it is obvious that $(C_5)$ is verified. For $d > 1$, we have $B_1 > d$ Taking logarithms and using Stirling's formula in the definition of $B_1$ and $B_2$, it is readily seen that $B$ has the announced asymptotic size. Since there is always a prime between $B$ and $2B$, the proposition follows. □

*4.3.2. Probabilistic strategies*

We begin with a "Monte Carlo"-like method, best described by the following algorithm. We need two auxiliary procedures : The function call `RandomPrime(A,C)` returns a random prime number in the real interval $[A, C]$. We assume that the numbers returned are uniformly distributed in the set of primes belonging to $[A, C]$ (see (Shoup (2005)), Section 7.5). The function `Nextprime` gives the smallest prime larger than the argument.

```
MCGoodPrime(F, M_γ, B', ε)
Input:
   F : A monic squarefree polynomial of degree d > 1 in K[X,Y].
   M_γ : A monic irreducible polynomial in Z[T].
   B' : A real number such that prime factors of N_F are less or equal to B'.
   ε : A real number with 0 < ε ≤ 1.
Output:
      A prime number p satisfying (C_5) with probability no less than 1 − ε.
Begin
   If  B' < 100 then Return NextPrime(B') End
   K ← 6 ln B'/(ε ln ln B') + 2d/ ln d
   C ← max {2d, K(ln K)^2}
   Return RandomPrime(d + 1, C)
End.
```

**Proposition 52.** MCGoodPrime($F$,$M_\gamma$,$B'$,$\epsilon$) returns a prime $p$ satisfying :

$$\mathrm{ht}(p) \in O(\log \log B' + \log d + \log \epsilon^{-1}).$$

In particular, if $B' = \max\{b, B_1, B_2\}$ (see (14) and (15)), then :

$$\mathrm{ht}(p) \in O(\log(dw \log n) + \log \mathrm{ht}(F) + \log \mathrm{ht}(M_\gamma) + \log \epsilon^{-1}).$$

Moreover, the probability that $p$ does not satisfy $(C_5)$ is less than $\epsilon$.

**Proof.** First note that condition $p > d$ is automatically verified. If $n$ is a positive integer, $\omega(n)$ is classically the number of prime numbers dividing $n$. For a positive real number $x$, we use the notation $\pi(x)$ for the number of primes less or equal to $x$. Estimates of (Bach and Shallit (1996), Section 8.8) give :

$$\frac{x}{\ln x} < \pi(x) \quad (x \geq 17), \qquad \pi(x) < \frac{2x}{\ln x} \quad (x > 1), \qquad \omega(n) < \frac{2 \ln n}{\ln \ln n} \quad (n \geq 100).$$

For $B' \geq 100$, the bound for $\omega$ yields :

$$\omega(N_F) \leq \omega(\mathrm{denom}(F)) + \omega(\mathrm{Norm}_{K/\mathbb{Q}}(R_F(T))) + \omega(\mathrm{Disc}(M_\gamma)) \leq 6 \frac{\ln B'}{\ln \ln B'}.$$

If $B' < 100$, the algorithm returns a correct result of size $O(1)$ with probability 1. We therefore assume in the sequel that $B' \geq 100$. Let $C$ be a number greater than $2d$ and $\tau$ be the probability that a prime given by RandomPrime($d+1$,$C$) divides $N$. We just have to determine a $C$ large enough so that $\tau \leq \epsilon$. But there is always a prime number between $d$ and $2d$ if $d > 1$, thus $\pi(C) - \pi(d) \geq 1$. Since RandomPrime has a uniform behavior, we search for a $C$ such that :

$$\tau = \frac{\omega(N_F)}{\pi(C) - \pi(d)} \leq \epsilon. \tag{16}$$

Since $B' \geq 100$ and $d > 1$, estimates above show that it is sufficient to find $C$ with :

$$\pi(C) \geq K = \frac{6 \ln B'}{\epsilon \ln \ln B'} + \frac{2d}{\ln d}.$$

Setting $C = K(\ln K)^2$, we find $C/\ln C = K(\ln K)^2/(\ln K + 2\ln\ln K)$. For $B' \geq 100$, $K$ is greater than 23 and therefore $(\ln K)^2/(\ln K + 2\ln\ln K) \geq 1$. Moreover $C \geq 226 > 17$, hence :
$$\pi(C) \geq \frac{C}{\ln C} \geq K,$$
and (16) holds. Moreover, the algorithm returns a prime $p$ with $\mathrm{ht}(p) = \max\{\log C, \log 2d\}$. Since $\log C = \log K + 2\log\log K \in O(\log\log B' + \log d + \log \epsilon^{-1})$, the asymptotic result follows. For the particular value of $B'$, the asymptotic bound can be deduced from Proposition 51. □

Finally, we consider a "Las Vegas" flavored method :

LVGoodPrime($F$,$M_\gamma$)
Input:
 $F$ : A monic squarefree polynomial of degree $d > 1$ in $K[X,Y]$.
 $M_\gamma$ : A monic irreducible polynomial in $\mathbb{Z}[T]$.
Output:
  A prime number $p$ satisfying $(C_5)$.
Begin
  $N_F \leftarrow \mathrm{denom}(F)|\mathrm{Norm}_{K/\mathbb{Q}}(R_F(T))\mathrm{Disc}(M_\gamma)|$
  $B' \leftarrow \max\{\mathrm{denom}(F), |\mathrm{Norm}_{K/\mathbb{Q}}(R_F(T))|, |\mathrm{Disc}(M_\gamma)|\}$
  Repeat
    $p \leftarrow$ MCGoodPrime($F, M_\gamma, B', 1/2$)
  until $p$ does not divide $N_F$ End
  Return $p$
End.

**Proposition 53.** LVGoodPrime($F$,$M_\gamma$) returns a prime $p$ satisfying :
$$\mathrm{ht}(p) \in O(\log{(dw\log n)} + \log\mathrm{ht}(F) + \log\mathrm{ht}(M_\gamma)).$$
and the average number of iterations is less than 2.

**Proof.** Obvious from Proposition 52. □

The computation of $B'$ and $N_F$ may have a significant cost. In our monodromy context, though, we need to determine $\Delta_F$ anyway. Moreover, in practice, we do not compute the norm of $\Delta_F$ trailing coefficient, but perform reduction modulo $\mathfrak{p} = (p, \overline{M})$ instead.

4.4. *Computation of $\mathcal{T}(F)$*

Let $K$ be an algebraic number field and let $\mathfrak{o}$ be the ring of algebraic integers of $K$. We assume that $F \in K[X,Y]$ and satisfies the notations and assumptions of section 1. If $\mathfrak{p}$ is a prime ideal of $\mathfrak{o}$ such that $F$ has good reduction at $\mathfrak{p}$, the computation of $\mathcal{T}(F)$ is straightforward :
- determine a finite field $\mathbb{F}_{p^t}$ isomorphic to $\mathfrak{o}/\mathfrak{p}$ and $\overline{F}$, image of $F$ under this isomorphism,
- compute $\mathcal{RT}(\mathbb{F}_{p^t}, \overline{F})$ using RNPuiseux,
- deduce $\mathcal{T}(\overline{F})$ using Proposition 30 (at the cost of a tree traversal),
- by Theorem 43, $\mathcal{T}(F) = \mathcal{T}(\overline{F})$.

## 5. Complexity of `RNPuiseux` over a finite field

In this section, $L$ denotes a finite field and $F$ belongs to $L[X, Y]$. Otherwise, we keep the notations and assumptions of Section 1. We denote by $p > d$ the characteristic of $L$. We also define $t_0 = [L : \mathbb{F}_p]$.

This section is devoted to the proof of the following theorem :

**Theorem 54.** *Assuming that FFT-based polynomial multiplication over finite fields is used, the `RNPuiseux` algorithm can compute the singular parts of a system rational Puiseux expansions above $0$ of $F$ in $\tilde{O}(d^3 n^2 + d^2 n t_0 \log p)$ field operations in $L$.*

As usual the notation $\tilde{O}$ hides logarithmic factors.

This result improves the bound of (Duval (1989)), which is in $O(d^6 n^2)$ field operations. Our estimates include factorization cost, while Duval relies on the D5 system to avoid factorizations. The gain comes from :
- truncation of powers of $X$ in the course of the algorithm (see Proposition 57),
- reducing transformations to shifts of univariate polynomials, for which fast methods are available (Proposition 61),
- a bound for $\delta_F$ (see below and Proposition 63).

For our application to the monodromy problem we need to consider expansions above *all* conjugacy classes over $L$ of critical points. More precisely, if $\Delta_F = \prod_i \Delta_i^{k_i}$ is a factorization of $\Delta_F$ into a product of irreducible polynomials of $L[X]$, rational Puiseux expansions above roots of $\Delta_i$ are conjugated over $L$. Therefore, it is sufficient to compute a system of rational Puiseux expansions above one root $c_i$ of $\Delta_i$ for each $i$. We obtain the following and remarkably similar theorem :

**Theorem 55.** *Assuming that FFT-based polynomial multiplication over finite fields is used, the `RNPuiseux` algorithm can compute the singular parts of systems of rational Puiseux expansions above all conjugacy classes over $L$ of critical points of $F$ in $\tilde{O}(d^3 n^2 t_0 \log p)$ field operations in $L$.*

We first introduce notations and make some assumptions :

- $\{R_i\}_{1 \leq i \leq \rho}$ with $R_i(T) = (X_i(T), Y_i(T))$ stands for singular parts of a system of rational Puiseux expansion above $0$,
- $(G_i, P_i, Q_i)$ is the output of `RNPuiseux` corresponding to $R_i$,
- $(r_i, e_i, f_i)$, $1 \leq i \leq \rho$ are respectively the regularity index, the ramification index and the coefficient field degree over $L$ of $R_i$.
- For each rational Puiseux expansion $R_i$, we can deduce $e_i f_i$ Puiseux series denoted $S_{ijk}(X), 1 \leq k \leq e_i, 1 \leq j \leq f_i$.
- $Y_{ijk}$ is the singular part of $S_{ijk}$.
- We define $\hat{F} = \prod_{ijk} Y - Y_{ijk}$.
- The following quantity will enter our estimates :

$$\delta_F = \sum_{i=1}^{\rho} f_i r_i.$$

- $L_t$ denotes an extension of degree $t$ of $L$.

- $M(N)$ will denote the number of field operations in $L_t$ needed to compute the product of two polynomials in $L_t[Z]$ of degree no larger than $N$. We recall that $M(N) \in O(N^2)$ for classical arithmetic and $M_t(N) \in O(N \log N \log \log N) \subset O^{\sim}(N)$ if FFT-based multiplication is used.
- It will be convenient to introduce $\tilde{M}(N) = M(N)/N$, so that $\tilde{M}(N) \in O(N)$ or $\tilde{M}(N) \in O(\log N \log \log N) \subset O^{\sim}(1)$.
- A field operation in $L_t$ can be made using $O(M(t) \log t)$ field operations in $L$.

We refer the reader to (von zur Gathen and Gerhard (1999)) for assertions regarding the complexity of operations over finite fields.

**Remark 56.** It is worth noting that $\delta_F$ is an upper bound for the number of elements of $L$ necessary to represent the $Y_i$. Indeed, each $Y_i$ has at most $r_i$ nonzero coefficients and each of those may be represented by at most $f_i$ elements of $L$. Moreover, assume that a dense representation is used for the truncated power series $Y_i$ (for instance, a vector of $r_i$ elements of $L_{f_i}$) and assume in turn that the coefficients of $Y_i$ are represented by vectors of $f_i$ elements of $L$. Then, $\delta_F$ is precisely equal to the size of the output.

We split the proof into several results.

*5.1. Truncating powers of $X$*

We prove the following proposition :

**Proposition 57.** *Systems of rational Puiseux expansions for $F$ and $\widetilde{F}^{\delta_F}$ above 0 have the same singular parts (up to trivial changes of the parameter $T$). Moreover, singular parts of rational Puiseux expansions of $F$ can be computed by applying* RNPuiseux *to $\widetilde{F}^{\delta_F}$ and truncating polynomials $H$ modulo $X^{\delta_F+1}$ at each stage of the algorithm.*

**Proof.** To obtain singular parts of rational Puiseux expansions of $F$, we just need $G_i$ modulo $X$ for all $i$. Setting $N_0 = 0$ in the bound of Lemma 58 below and applying Lemma 59, we see that computations modulo $\delta_F + 1$ are sufficient. □

To simplify notations in the next lemma, we drop indices : $R$ is a rational Puiseux expansion and $(r, e, f)$ and $(G, P, Q)$ are the associated quantities. Let $m_k i + q_k j = l_k$, $(1 \leq k \leq h)$ the sequence of Newton polygon edges encountered in RNPuiseux while computing $(G, P, Q)$, and $F = H_0, H_1, \ldots, H_h = G$ be the sequence of input polynomials.

**Lemma 58.** *Let $N_0$ be a positive integer. In order to compute $\widetilde{G}^{N_0}$, it is sufficient, at each stage of the algorithm, to compute $H_k$ modulo $X^{N+1}$, where $N = \frac{N_0}{e} + \sum_{k=1}^{h} \frac{l_k}{q_1 \ldots q_k}$.*

**Proof.** The algorithm performs a sequence of substitution :
$$H_{k+1}(X, Y) = \frac{H_k(\xi^{v_k} X^{q_k}, X^{m_k}(\xi^{u_k} + Y))}{X^{l_k}}.$$
If $H_k(X, Y) = \sum_{ij} \alpha_{ij} X^j Y^i$, we define $H_{kw}(X, Y) = \sum_{m_k i + q_k j = w} \alpha_{ij} X^j Y^i$, so that $H_k = \sum_w H_{kw}$. Monomials of $H_{kw}$ are transformed into monomials of $H_{k+1}$ arranged on the horizontal line $j = w - l_k$ (see Figure 5.1). Therefore, to determine $\widetilde{G}^{N_0} = \widetilde{H}_h^{N_0}$ it is sufficient to know $\widetilde{H}_{h-1}^{(N_0+l_h)/q_h}$. Proceeding recursively, we obtain the given value of $N$ and conclude that it is sufficient to compute all $H_k$ modulo $X^{N+1}$. □

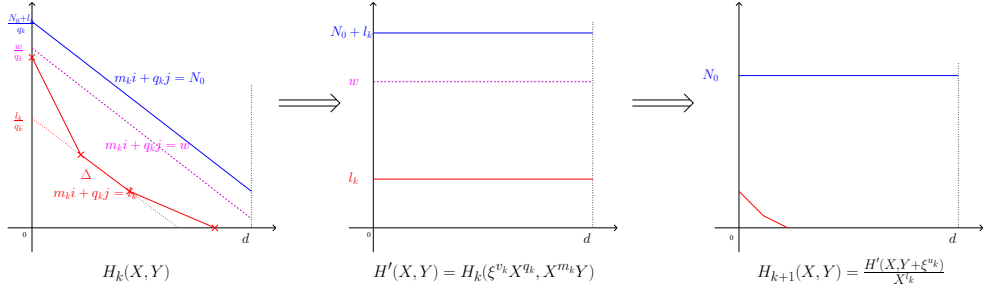

Fig. 4. Geometric interpretation of a substitution

This lemma may also be useful if one wishes to compute expansions beyond the regularity index. It is clear from the proof that we could adopt an adaptative strategy and truncate further $H_k$ by taking into account the information gained at the current stage of the execution. It is unlikely that this approach will improve the asymptotic complexity.

**Lemma 59.** *With the notation of Lemma 58 :* $\sum_{k=1}^{h} \frac{l_k}{q_1 \ldots q_k} \leq \delta_F$.

**Proof.** Since $F$ and $\hat{F}$ have the same rational Puiseux expansion singular parts, applying RNPuiseux to $\hat{F}$ yields the same sequence $(m_k, q_k, l_k)_{1 \leq k \leq h}$ as for $F$ and an output $(\hat{G}, P, Q)$, where $P(X) = \lambda X^e$ and $Q(X, Y) = Q_0(X) + X^r Y$ are the polynomials associated to $F$. Set $c = l_h + l_{h-1} q_h + \cdots + l_1 q_2 \ldots q_h$. From the definition of the substitutions, we get :
$$\hat{G}(X, Y) = \frac{\hat{F}(P(X), Q(X, Y))}{X^c}.$$
Since $(1, 0)$ belongs to $\mathcal{GN}(\hat{G})$, we have $v_X(\hat{G}_Y(X, 0)) = 0$. Hence, Taylor formula gives $c = v_X(X^r \hat{F}_Y(P(X), Q_0(X)))$, which is equivalent to :
$$\frac{c}{e} = \frac{r + v_X(\hat{F}(P(X), Q_0(X)))}{e} = \frac{r}{e} + v_X(\hat{F}_Y(X, Q_0((X/\lambda)^{1/e}))).$$
Assume that the truncated Puiseux series $Q_0((X/\lambda)^{1/e})$ is equal to $Y_{i'j'k'}$. Then :
$$\hat{F}_Y(X, Q_0((X/\lambda)^{1/e})) = \hat{F}_Y(X, Y_{i'j'k'}) = \prod_{\substack{(i,j,k) \\ (i,j,k) \neq (i',j',k')}} (Y_{i'j'k'} - Y_{ijk}).$$
Now, since $v_X(Y_{i'j'k'} - Y_{ijk}) \leq \frac{r_i}{e_i}$, we have :
$$\frac{c}{e} \leq \frac{r_{i'}}{e_{i'}} + \sum_{\substack{(i,j,k) \\ (i,j,k) \neq (i',j',k')}} \frac{r_i}{e_i} \leq \sum_{i=1}^{\rho} r_i f_i.$$
□

## 5.2. Cost of substitutions

**Lemma 60.** *Let $N$ be a positive integer. Let $H \in L_t[X,Y]$, $\xi \in L_t$, and define $H'(X,Y) = H(\xi^v X^q, X^m(\xi^u + Y))/X^l$, where $(m,q,l)$ corresponds to an edge of $\mathcal{GN}(H)$. Then, we can compute the coefficients of $\widetilde{H'}^N$ with $O(NM(d))$ field operations in $L_t$.*

**Proof.** Write $H(X,Y) = \sum_{w=l}^{N+l} H_w(X,Y)$ with $H_w(x,y) = \sum_{mi+qj=w} \alpha_{ij} X^j Y^i$.

$$H_w(\xi^v X^q, \xi^u X^m(1+Y)) = \sum_{mi+qj=w} \alpha_{ij}(\xi^v X^q)^j (X^m(\xi^u + Y))^i$$
$$= X^w \sum_{mi+qj=k} \alpha_{ij} \xi^{vj}(\xi^u + Y)^i$$
$$= X^w R_w(Y + \xi^u)$$

where $R_w(Z) = \sum_{mi+qj=w} \alpha_{ij} \xi^{vj} Z^i$ is a univariate polynomial of degree at most $d$.

We first form the necessary powers of $\xi^v$: The exponent is bounded by $(N+l)/q$ but $l/q < d$ since slopes of generic Newton polygons are at least -1. Computing all powers up to this bound is achieved in $O(N+d)$ operations in $L_t$. Constructing the polynomials $R_w$ requires at most $Nd$ multiplications in $L_t$. Finally, since $p > d$, the shift in $R_w$ can be reduced essentially to the multiplication of two degree $d$ polynomials, with cost $O(M(d))$ (Bini and Pan (1994)). Since we must perform $N$ such shifts, the total cost is in $O(M(d)N)$. $\square$

**Proposition 61.** *The substitutions needed to compute the singular parts of a system of rational Puiseux expansion of $F$ requires $O\left(\delta_F^2 M(d) \tilde{M}(d) \log d\right)$ field operations in $L$.*

**Proof.** To compute a rational Puiseux expansion $R_i$, `RNPuiseux` performs at most $r_i$ substitutions over the field $L_{f_i}$. Proposition 57 ensures that all substitutions can be done modulo $X^{\delta_F+1}$. From Lemma 60, substitutions for computing $R_i$ cost at most $O(M(d)\delta_F r_i)$ operations in $L_{f_i}$. Taking into account the extension $L_{f_i}/L$ and summing over $i$, we obtain a total cost:

$$O\left(M(d)\delta_F \sum_{i=1}^{\rho} r_i M(f_i) \log f_i\right) = O\left(M(d)\delta_F \sum_{i=1}^{\rho} r_i f_i \tilde{M}(f_i) \log f_i\right)$$
$$= O\left(\delta_F^2 M(d) \tilde{M}(d) \log d\right).$$

$\square$

### 5.3. Factorizations cost

**Proposition 62.** *All factorizations of characteristic polynomials required by* `RNPuiseux` *can be computed with an expected number of field operations in $L$ :*

$$O\left(\delta_F \log d \left[M(d^2) + t_0 \log p M(d) \log d\right]\right)$$

**Proof.** Let $L_t$ be the extension of $L$ over which a factorization of $\phi_\Delta$ must be determined. It is easily seen that degree $d_\Delta$ of $\phi_\Delta$ is at most $d/t$. This is obviously true at the first stage of the algorithm, where $t = 1$. Assume that the property is true at a given stage of the algorithm and denote by $\xi$ a root of multiplicity $k$ of $\phi_\Delta$. If $\phi_{\Delta'}$ is a characteristic polynomial of the polygon induced by the choice of $\xi$ and $d_{\Delta'}$ is its degree, by Proposition 25 :

$$d_{\Delta'} \leq k \quad \text{and} \quad k[L_t(\xi) : L_t] \leq d_\Delta \leq d/t.$$

Hence, $d_{\Delta'} \leq d/[L_t(\xi) : L]$ and the property is true by induction. Hence, factorization of a characteristic polynomial can be achieved with an average number of $O(M((\frac{d}{t})^2) \log \frac{d}{t} + t_0 t \log p M(\frac{d}{t}) \log \frac{d}{t})$ operations in $L_t$ (von zur Gathen and Gerhard (1999), Corollary 14.30). Each operation in $L_t$ can be performed with $O(M(t) \log t)$ operations in $L$. It is easily verified that $O(M((\frac{d}{t})^2) \log \frac{d}{t} M(t) \log t) \subset O(M(d^2) \log d)$ and $O(M(\frac{d}{t}) \log \frac{d}{t}) \subset O(M(d) \log^2 d)$, no matter which arithmetic is used. Hence, we obtain :

$$O\left(M(d^2) \log d + t_0 t \log p M(d) \log^2 d\right).$$

Finally, we multiply by $r_i$, bound $t$ by $f_i$ and summing over $i$ to obtain the result. □

### 5.4. Bounding $\delta_F$

**Proposition 63.**

$$\delta_F = \sum_{i=1}^{\rho} r_i f_i \leq v_X(\Delta_F)$$

**Proof.** By definition of the regularity index, for each Puiseux series $S_{ijk}(X)$, there exists an other Puiseux series $S_{i_0 j_0 k_0}(X)$ with $i_0 \in \{1 \ldots \rho\}$, $j_0 \in \{1 \ldots f_{i_0}\}$ and $k_0 \in \{1 \ldots e_{i_0}\}$ such that $\frac{r_i-1}{e_i} < v_X(S_{ijk}(X) - S_{i_0 j_0 k_0}(X)) \leq \frac{r_i}{e_i}$. Moreover, if $v_X(S_{ijk}(X) - S_{i_0 j_0 k_0}(X)) \neq \frac{r_i}{e_i}$, $e_i$ is a proper divisor of $e_{i_0}$. Thus, denoting $q = e_{i_0}/e_i > 1$, there exists $m \in \mathbb{N}$ and $\alpha \neq 0 \in \overline{L}$ such that $1 \leq m < q$ and :

$$S_{i_0 j_0 k_0}(X) = \widetilde{S_{ijk}}^{\frac{r_i}{e_i}}(X) + \alpha X^{\frac{r_i-1}{e_i} + \frac{m}{e_{i_0}}} + \cdots$$

Hence, for $0 \leq l \leq q-1$, we have :

$$S_{i_0 j_0 k_0}^{[l e_i, e_{i_0}]}(X) = \widetilde{S_{ijk}}^{\frac{r_i}{e_i}}(X) + \zeta_q^{ml} \alpha X^{\frac{r_i-1}{e_i} + \frac{m}{e_{i_0}}} + \cdots$$

Therefore, we obtain :

$$\sum_{l=0}^{q-1} v_X(S_{ijk}(X) - S_{i_0 j_0 k_0}^{[l e_i, e_{i_0}]}(X)) = q \frac{r_i}{e_i} + \frac{q-m}{e_i} \geq \frac{r_i}{e_i}$$

For each Puiseux series $S_{ijk}$, we have :

$$v_X(F_Y(X, S_{ijk}(X))) = \sum_{\substack{(i',j',k') \\ (i',j',k') \neq (i,j,k)}} v_X(S_{ijk}(X) - S_{i'j'k'}(X)) \geq \frac{r_i}{e_i}.$$

Summing over $(i, j, k)$, relation (1) gives :

$$v_X(\Delta_F) = \sum_{(i,j,k)} v_X(F_Y(X, S_{ijk}(X))) \geq \sum_{i=1}^{\rho} \sum_{j=1}^{f_i} \sum_{k=1}^{e_i} \frac{r_i}{e_i} = \sum_{i=1}^{\rho} r_i f_i.$$

□

5.5. *Proof of Theorem 54 and 55*

It is interesting to bound first the number of operations in $L$ in terms of the output size, namely $\delta_F$.

**Theorem 64.** *The number of operations in $L$ required to compute singular parts of rational Puiseux expansions of $F$ above 0 is in :*

$$\tilde{O}\left(\delta_F M(d) [\delta_F + M(d) + t_0 \log p]\right) \subset \tilde{O}\left(n\, M(d) d\, [nd + M(d) + t_0 \log p]\right).$$

**Proof.** The first assertion follows from Proposition 61 and Proposition 62 since $M(d^2) \in O(M(d)^2)$ and $\tilde{M}(d) \log d \in \tilde{O}(1)$. The inclusion is a consequence of Proposition 63, where $v_X(\Delta_F)$ is bounded from above by $\deg \Delta_X \leq (2d-1)n$. □

Theorem 54 is now a trivial consequence of this result.

As for Theorem 55, we proceed as follow :

First of all, the calculation of the discriminant $\Delta_F$ can be done in $O(nM(nd) \log(nd))$ field operations in $L$ (von zur Gathen and Gerhard (1999)). Thus, this step is included in our complexity bound.

Moreover, $\Delta_F$ is a polynomial in $X$ with degree $\delta$ at most $(2d-1)n$. Therefore, it can be factorized over $L$ in $O(M(\delta^2) \log \delta + t_0 \log p M(\delta) \log \delta) \subset \tilde{O}(n^2 d^2 + ndt_0 \log p)$ field operations in $L$ using fast multiplication (von zur Gathen and Gerhard (1999)). Hence, this step is also included in our complexity bound.

Then, if $\Delta_F = \prod_{i=1}^{m} \Delta_i^{k_i}$ is the factorization obtained, set $t_i = \deg_X(\Delta_i)$ and let $c_i$ be a root of $\Delta_i$. The computation of $F_i = F(X + c_i, Y)$ can be performed at the cost of $d$ shifts in $L(c_i) = L_{t_i}$ of the coefficients of $F$ in $Y$. Hence, the complexity of this step is in $O(dM(n))$ field operations in $L_{t_i}$ (Bini and Pan (1994)) and $O(dM(n)M(t_i) \log t_i) \subset \tilde{O}(dnt_i)$ operations in $L$. Summing over $i$ and bounding $\sum_i t_i$ by $(2d-1)n$ gives again a correct estimate.

Finally, we note that the quantity $\delta_{F_i}$ associated to $F_i$ is bounded by $k_i$ (see Proposition 63). From Theorem 64, using fast multiplication, the function call $\texttt{RNPuiseux}(F(X + c_i, Y), L(c_i))$ requires $\tilde{O}(k_i d[k_i + d + t_0 t_i \log p])$ field operations in $L_{t_i}$, and so $\tilde{O}(k_i t_i d[k_i + d + t_0 t_i \log p]) \subset \tilde{O}(k_i t_i n d^2 t_0 \log p)$ field operations in $L$. Summing over $i$ allows to conclude since $\sum_i k_i t_i = \delta \leq (2d-1)n$.

## 6. Bit-complexity

Let $F$ be a polynomial of $K[X,Y]$, where $K$ is an algebraic number field represented as in Section 4.3. We recall that $[K : \mathbb{Q}] = w$. We study the bit-complexity of the computation of $\mathcal{T}(F)$. We estimate only word operations generated by arithmetic operations in various coefficient fields. Assuming some care is taken in the implementation, (for instance, access to coefficients of polynomial should be achieved in constant time), results below should give a realistic upper bound for the behavior of an actual program.

Our bounds for randomized algorithms do not include the cost of generating prime numbers, nor the cost of computing bounds given by our formula.

We assume that elements of $\mathbb{F}_p$ are represented by nonnegative integers. In order to simplify expressions, we assume that fast arithmetic is used for integer arithmetic as well as polynomial arithmetic over finite fields.

**Theorem 65.** *Given a real number $\epsilon$ with $0 < \epsilon \leq 1$, there exists a probabilistic Monte Carlo algorithm that computes $\mathcal{T}(F)$ with an expected number of :*

$$\widetilde{O}(d^3 n^2 w^2 \operatorname{ht}(p)^2 [\operatorname{ht}(M_\gamma) + \operatorname{ht}(F)]) \subset \widetilde{O}(d^3 n^2 w^2 \log^2 \epsilon^{-1} [\operatorname{ht}(M_\gamma) + \operatorname{ht}(F)])$$

*word operations and probability of error less than $\epsilon$.*

**Proof.** Call `MCGoodPrime`($F,M_\gamma,B',\epsilon$) with $B' = \max\{b, B_1, B_2\}$ to get a good prime $p$ with probability of error less than $\epsilon$ and size $\operatorname{ht}(p)$ given by Proposition 52. Reduce the coefficients of $M_\gamma$ and $F_n$ modulo $p$, and check that $p$ does not divide $b = \operatorname{denom}(F)$. Denote $\tilde{F}_n$ the reduction of $F_n$. This step requires $\widetilde{O}(w\operatorname{ht}(M_\gamma) + wnd\operatorname{ht}(F_n) + \operatorname{ht}(b))\log p$ word operations. Then, factorize $\overline{M_\gamma}$ over $\mathbb{F}_p$ at the cost of $\widetilde{O}(w^2 + w\log p)$ operations in $\mathbb{F}_p$ and choose an irreducible factor $\overline{M}$ of minimal degree. Coefficients of $\tilde{F}_n$ must be reduced modulo $\overline{M}$ and $p$. There are $nd$ coefficients and each division induces $\widetilde{O}(w)$ operation in $\mathbb{F}_p$. The worst case for `RNPuiseux` occurs if $\overline{M}$ happens to have degree $w$. In this case, call `RNPuiseux`($\mathbb{F}_{p^w}, F$), which require $\widetilde{O}(d^3 n^2 + d^2 nw \log p)$ operations in $\mathbb{F}_{p^w}$ by Theorem 54, each operation requiring $\widetilde{O}(w)$ operations in $\mathbb{F}_p$. In summary, $\widetilde{O}(d^2 nw[dn + w\log p + \operatorname{ht}(M_\gamma) + \operatorname{ht}(F_n)]\log p)$ word operations in are needed. Taking into account the bound for $\operatorname{ht}(p) = \log p$ and applying crude estimates yield the result, since $\log \operatorname{ht}(F)$ and $\log \operatorname{ht}(M_\gamma)$ are logarithmic terms in the size of the input. □

**Remark 66.** It is possible to demonstrate without much difficulty that the same bit-complexity result holds for the computation of polygon trees of $F$ above all critical points. In turn, this result gives a bit-complexity bound for the computation of the genus of an algebraic curve, a problem for which we have not found similar results in the literature.

## 7. Conclusion

This paper summarizes the results we have obtained regarding the symbolic part of our program towards a fast and reliable method to compute Puiseux series with floating point coefficients. In particular, the criterion ensuring preservation of polygon trees is essential.

Along the path, we have derived improved complexity bounds for the computation of Puiseux series over finite fields. Although not optimal, these bounds are quite reasonable, i.e. quadratic in the output size, up to a factor $d$. Refinements may be obtained by optimizing several steps of our complexity analysis. It is not clear that these refinements may significantly improve the final asymptotic cost.

Bit-complexity estimates for the Monte-Carlo version of the first stage of our symbolic-numeric method confirm that the coefficient growth of pure symbolic Newton-Puiseux algorithm is avoided. Complexity bounds for the Las Vegas and deterministic versions can be obtained similarly. An analogous bit-complexity bound can also be deduced for the computation of the genus of an algebraic curve defined over an algebraic number field.

The numerical part of the algorithm is still being investigated and several variants are being compared.

The scope of our symbolic-numeric strategy may be broader : For instance, series solutions at singular points of certain linear differential equations can be constructed using Newton polygons. It would be interesting to investigate the benefits of our approach in this context.

## Acknowledgements

The authors would like to thank Mark van Hoeij for stimulating conversations. We are also grateful to an anonymous referee of (Poteaux (2007)) for valuable suggestions.